\documentclass[a4paper,11pt]{article}
\usepackage{amssymb,graphicx,amsmath}

\renewcommand{\theequation}{\thesection.\arabic{equation}}
\input{tcilatex}

\newcommand{\bc}{\begin{center}}
\newcommand{\ec}{\end{center}}
\def\ba#1{\begin{array}{#1}\displaystyle}
\newcommand{\ea}{\end{array}}
\newcommand{\z}{\\[2mm] \displaystyle}

\newcommand{\beq}{\begin{equation}}
\newcommand{\eeq}{\end{equation}}
\newcommand{\beqa}{\begin{eqnarray}}
\newcommand{\eeqa}{\end{eqnarray}}
\newcommand{\no}{\nonumber}
\newcommand{\n}{\nonumber\\}
\newcommand{\bi}{\begin{itemize}}
\newcommand{\ei}{\end{itemize}}

\def\sect#1{\section{#1}\setcounter{equation}{0}}

\def\lt#1{\left#1}
\def\rt#1{\right#1}
\def\t#1{\tilde{#1}}
\def\h#1{\hat{#1}}
\def\b#1{\bar{#1}}
\def\frc#1#2{\frac{#1}{#2}}

\newcommand{\p}{\partial}

\newcommand{\bra}{\langle}
\newcommand{\ket}{\rangle}
\newcommand{\Z}{{\mathbb{Z}}}
\newcommand{\N}{{\mathbb{N}}}
\newcommand{\R}{{\mathbb{R}}}

\newcommand{\Or}{{\cal O}}

\newcommand{\ep}{\epsilon}
\newcommand{\varep}{\varepsilon}

\newcommand{\Tr}{{\rm Tr}}

\newcommand{\tw}{{\cal T}}
\newcommand{\orb}{{\cal M}}
\newcommand{\sym}{\sigma}

\newcommand{\rx}{{\rm x}}
\newcommand{\ry}{{\rm y}}

\begin{document}

\setcounter{page}{0} \topmargin0pt \oddsidemargin0mm \renewcommand{%
\thefootnote}{\fnsymbol{footnote}} \newpage \setcounter{page}{0}
\begin{titlepage}

\vspace{0.2cm}
\begin{center}
{\Large {\bf Bi-partite entanglement entropy in massive
1+1-dimensional quantum field theories}}

\vspace{0.8cm} {\large  \text{Olalla
A.~Castro-Alvaredo$^{\bullet}$ and Benjamin Doyon$^{\circ}$}}

\vspace{0.2cm}{$^{\bullet}$  Centre for Mathematical Science, City University London, \\
Northampton Square, London EC1V 0HB, UK} \\
\vspace{0.2cm} {$^{\circ}$  Department of Mathematical Sciences,
Durham University \\ South Road, Durham DH1 3LE, UK }
\end{center}
\vspace{1cm}

\noindent This manuscript is a review of the main results obtained
in a series of papers involving the present authors and their
collaborator J.L.~Cardy over the last two years. In our work we
have developed and applied a new approach for the computation of
the bi-partite entanglement entropy in massive 1+1-dimensional
quantum field theories. In most of our work we have considered
these theories to be also integrable. Our approach combines two
main ingredients: the ``replica trick," and form factors for
integrable models and more generally for massive quantum field
theory. Our basic idea for combining fruitfully these two
ingredients is that of the branch-point twist field. By the
replica trick we obtained an alternative way of expressing the
entanglement entropy as a function of the correlation functions of
branch-point twist fields. On the other hand, a generalisation of
the form factor program has allowed us to study, and in integrable
cases to obtain exact expressions for, form factors of such twist
fields. By the usual decomposition of correlation functions in an
infinite series involving form factors, we obtained exact results
for the infrared behaviours of the bi-partite entanglement
entropy, and studied both its infrared and ultraviolet behaviours
for different kinds of models: with and without boundaries and
backscattering, at and out of integrability.

\vfill{ \hspace*{-9mm}
\begin{tabular}{l}
\rule{6 cm}{0.05 mm}\\
$^\bullet \text{o.castro-alvaredo@city.ac.uk}$\\
$^\circ \text{benjamin.doyon@durham.ac.uk}$\\
\end{tabular}}

\renewcommand{\thefootnote}{\arabic{footnote}}
\setcounter{footnote}{0}
\end{titlepage}
\newpage

\tableofcontents

\sect{Introduction}

Entanglement is a fundamental property of quantum systems. Its most striking consequence is the fact that performing a local measurement may affect
the outcome of local measurements far away; this is
one of the main differences between quantum and classical systems.
Given that our
understanding of the physical world is largely based on everyday experience
and that entanglement seems to contradict such experiences, it is
not surprising that the existence of quantum entanglement has been
a source of controversy and heated scientific debate for some time
(see e.g. \cite{EPR} for the famous EPR-paradox). Today
entanglement is a well established and measurable phenomenon,
whose reality was experimentally confirmed in the early eighties
by the famous experiments of Alain Aspect and collaborators
\cite{aspect1,aspect2}, using pairs of maximally
entangled photons. In the last decades, many
applications of entanglement have developed into successful fields
of research such as quantum computation and quantum cryptography.
Entanglement lies also at the heart of other interesting phenomena
such as quantum teleportation (see \cite{quantuminfo} for a
review).

As a consequence of the prominent role of entanglement in quantum
physics, there has been great interest in developing efficient
(theoretical) measures of entanglement. For example, a quantity of
current interest in quantum models with many local degrees of
freedom is the bi-partite entanglement entropy \cite{bennet},
which we will consider in this review. In its most general
understanding, it is a measure of the amount of quantum
entanglement, in some pure quantum state, between the degrees of
freedom associated to two sets of independent observables whose
union is complete on the Hilbert space.  In the cases considered
in this review, the quantum state will mostly be the ground state
$|{\rm gs}\ket$ of some extended 1+1-dimensional (1 space + 1 time
dimension) quantum model, and the two sets of observables
correspond to the local observables in two connected regions, say
$A$ and its complement, $\bar{A}$ (we will also briefly discuss
the general technique in the case of excited states). Other
measures of entanglement exist, see e.g.
\cite{bennet,Osterloh,Osborne,Barnum,Verstraete}, which occur in
the context of quantum computing, for instance. Measures of
entanglement are important at a theoretical level, as they give a
good description of the quantum nature of a ground state, perhaps
more so than correlation functions.

General aspects of the entanglement entropy in extended quantum systems will be discussed
in \cite{CCD}. For our present purposes,
prominent examples of extended one-dimensional quantum systems are quantum spin chains, which model physical systems
consisting of infinitely long one-dimensional arrays of equidistant atoms
characterized by their spin. Their entanglement has been
extensively studied in the literature
\cite{Eisert,Latorre1,Latorre2,Latorre3,Jin,Lambert,Casini,KeatingM05,Weston}.
\begin{figure}[h!]
\begin{center}
 \includegraphics[width=9cm,height=4cm]{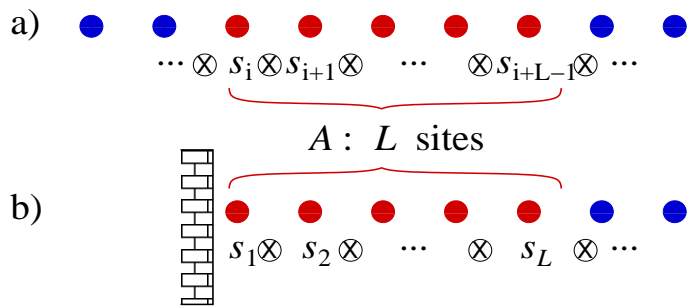}\\
  \caption{The entangled region $A$ of length $L$ of a quantum spin chain. a) the bulk case, b) the boundary case.}\label{chain}
  \end{center}
\end{figure}
\newline \indent In order to provide a formal definition of the
entanglement entropy, let us consider the Hilbert space of a
quantum model, such as the chain above, as a tensor product of
local Hilbert spaces associated to its sites. This can be written as a
tensor product of the two Hilbert spaces associated to the regions
$A$ and $\bar{A}$: \begin{equation} \label{Hdecomp} {\cal H} = {\cal A} \otimes
\b{{\cal A}}.
\end{equation} Then the entanglement entropy is the von
Neumann entropy of the reduced density matrix $\rho_A$ associated
to $A$: \beq\label{defeegs}
    S_A = -\Tr_{{\cal A}} \rho_A\log \rho_A ~,\quad \rho_A = \Tr_{\b{{\cal A}}} |{\rm gs}\ket \bra {\rm gs}|.
\eeq

We will be interested in analysing the entanglement entropy in the scaling limit of infinite-length quantum chains. The scaling limit gives the universal part of the quantum chain behaviour near a quantum critical point, which is
described by a model of 1+1-dimensional quantum field theory (QFT) (which we will assume throughout
to possess Poincar\'e invariance). The scaling limit is obtained
by approaching the critical point while letting the length $L$ of the
region $A$ go to infinity in a fixed proportion with the
correlation length $\xi$ (these lengths are measured in number of lattice sites). In this limit, the entanglement entropy is in fact divergent. The way the entanglement entropy diverges was understood in \cite{CallanW94,HolzheyLW94,Calabrese:2004eu}. It is controlled by the central charge $c$ corresponding to the critical point that is being approached. In general, for every point that separates a connected component of $A$ from a connected component of its complement $\b{A}$ (a boundary point of $A$), there is a term $c/6 \, \log\xi$. We will look at two cases: where the quantum chain is infinite in both directions and $A$ is a connected segment, with two  boundary points (the bulk case), and where the quantum chain is infinite only in one direction, and $A$ is a connected segment starting at the boundary of the chain, with only one true boundary point (the boundary case) -- see Fig. \ref{chain}. In these two cases, the divergent part of the entanglement entropy is respectively:
\beq\label{divSA}
    S_A^{\text{bulk}} \sim \frc{c}3 \log\xi + O(1),\quad S_A^{\text{boundary}} \sim \frc{c}6 \log\xi + O(1).
\eeq
These divergent terms are not universal (as they depend on the correlation length); the universal terms of the entanglement entropy are hidden in the $O(1)$ part. These universal terms, that depend on the proportion to $\xi$ with which $A$ is sent to infinity, are in general described by QFT (conformal, massive, etc.). The analysis of the universal terms using CFT techniques in various situations is reviewed in \cite{CalabreseCardyRev}. In the present review, we will explain how to use massive QFT techniques in order to obtain information about the universal terms. We will now overview the main ideas and results of our works on this subject.

\subsection{Overview of ideas and results}

\begin{figure}[h!]
\begin{center}
 \includegraphics[width=8cm,height=5cm]{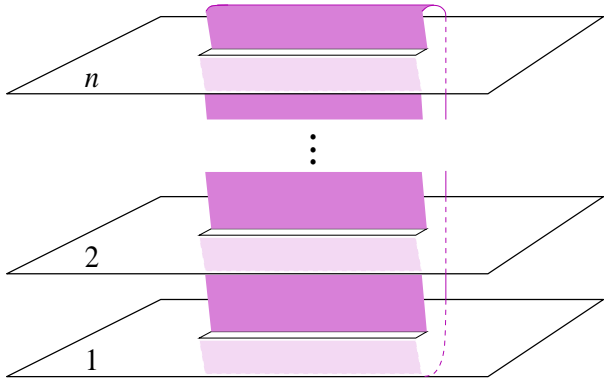}\\
  \caption{A representation of the Riemann
surface $\orb_{n,x_1,x_2}$.}\label{fig-sheets}
  \end{center}
\end{figure}

It is known since some time
\cite{CallanW94,HolzheyLW94,Calabrese:2004eu,Calabrese:2005in}
that the bi-partite entanglement entropy in the scaling limit can be re-written in
terms of more geometric quantities, using a method
known as the ``replica trick''. The essence of the method is to
``replace" the original QFT model by a new model consisting of
$n$ copies (replicas) of the original one, in order to use the formula
\beq\label{formulan1}
    S_A = -\lim_{n\to 1}\frc{d}{dn} \Tr_{{\cal A}} \rho_A^n.
\eeq The trace in this formula is reproduced by the condition that
these copies  be connected cyclicly through a finite cut on the
region $A$. Then, this trace is the partition function
$Z_n(x_1,x_2)$ of the original (euclidean) QFT model on a Riemann
surface ${\cal M}_{n,x_1,x_2}$ with two branch points, at the
points $x_1$ and $x_2$ in $\R^2$, and $n$ sheets cyclicly
connected (we will provide more explanations about this in the
next section). Fig.~\ref{fig-sheets} shows a representation of such a Riemann surface.
The positions of the branch points correspond to the end-points of
the region $A$ in the scaling limit. This gives: \beq\label{iden}
    S_A(|x_1-x_2|) = -\lim_{n\to 1}\frc{d}{dn} \frc{Z_n(x_1,x_2)}{Z_1^n}.
\eeq Here, $|x_1-x_2|$ is the euclidean distance between $x_1$ and
$x_2$. This formula holds both for our bulk and boundary cases. In
the boundary case, one of the branch points (say $x_1$) is on the
boundary of the model, $x_1=(0,0)$. The positions $x_1$ and $x_2$
are dimensionful positions in the QFT model. They are naturally at
zero imaginary time $x_1=(\rx_1,0)$, $x_2 = (\rx_2,0)$ (but this
is not crucial because the euclidean QFT has rotation invariance),
and their $\rx$-coordinates are related to the ratio between the
dimensionless region length $L$ and correlation length $\xi$ of
the quantum chain by $|\rx_1-\rx_2| = L/(m\xi)$, where $m$ is the
QFT mass scale associated to $\xi$ (whose only role here is to
provide a dimension).

Naturally, this expression implies that we must analytically
continue the quantity $Z_n(x_1,x_2)$ from $n\in \N$, where it is
naturally associated to Riemann surfaces, to $n\in[1,\infty)$. The
object $\Tr_{{\cal A}} \rho_A^n$ certainly has a well-defined
meaning for any $n$ such that ${\rm Re}(n)>0$. Indeed, $\rho_A$ is
hermitian (and has non-negative eigenvalues summing to 1), so that
$\Tr_{{\cal A}} \rho_A^n$ is the sum of the $n^{\rm th}$ powers of
its eigenvalues (with multiplicities). Note that this is an
analytic continuation from positive integers $n$ to complex $n$
that satisfies the requirements of Carlson's theorem
\cite{Rubel55}, hence the unique one that does. The scaling limit
of this object is what defines the proper analytic continuation of
$Z_n(x_1,x_2)$. Finding the correct analytic continuation has been
one of the major challenges encountered in our work. It is natural
to assume, as it has been done before \cite{CallanW94} and
discussed  in \cite{other} in the present context, that the two
branch points just become conical singularities with angle $2\pi
n$, the rest of the space being flat.

As will be described later in more detail, in \cite{entropy} we
showed that there is a way of associating the branch points at $x_1$ and
$x_2$ to local QFT fields: through {\em branch-point twist fields}
${\mathcal{T}}(x_1), \tilde{\mathcal{T}}(x_2)$. These twist fields
are defined only in the replica model (not in the original model),
and are associated to certain elements of the extra permutation
symmetry present in the replica model. In terms of these fields we
showed that, in the bulk case,
\begin{equation}
\frac{Z_n(x_1,x_2)}{Z_1^n}=\mathcal{Z}_n \varepsilon^{2d_n}
\langle 0|{\mathcal{T}}(x_1) \tilde{\mathcal{T}}(x_2)|0\rangle.
\label{result}
\end{equation}
Here $\langle 0|\cdots|0\rangle$ denote correlation functions in
the $n$-copy model; the state $|0\rangle$ is the vacuum state of
the latter. The branch-point twist fields have the CFT
normalisation  (which we will discuss later). The constant ${\cal
Z}_n$, with ${\cal Z}_1=1$, is an $n$-dependent non-universal
constant, $\varepsilon$ is a short-distance cut-off which is
scaled in such a way that $d {\cal Z}_n/dn=0$ at $n=1$, and,
finally, $d_n$ is the scaling dimension of the counter parts of
the fields $\mathcal{T}, \tilde{\mathcal{T}}$ in the underlying
$n$-copy conformal field theory,
\begin{equation}\label{dn}
    d_n=\frac{c}{12}\left(n-\frac{1}{n}\right),
\end{equation}
which can be obtained by CFT arguments
\cite{Calabrese:2005in,entropy} and where $c$ is the central
charge. The short-distance cut-off is related to the correlation
length  via $\varep = a /(m\xi)$ for some dimensionless finite
non-universal number $a$. This short distance cut-off takes care
of the infinite contributions to the partition functions
$Z_n(x_1,x_2)$, as compared to the $n^{\rm th}$ power of $Z_1$,
around the points $x_1$ and $x_2$ where a branch point lies. There
is one contribution of $\varep^d_n$ for each point, corresponding
to one contribution of $c/6 \log\xi$ to the entanglement entropy.

In our most recent work \cite{nexttonext} we have generalized this
understanding to the boundary case, where now the region $A$
extends between the origin $\rx_1=0$, where the boundary of the
model is located, and the $\rx$-coordinate $\rx_2$. With similar
arguments we have showed that
\begin{equation}
\frac{Z_n(0,x_2)}{Z_1^n}=\mathcal{Z}_n \varepsilon^{d_n}
\langle 0|{\mathcal{T}}(x_2)|B\rangle. \label{result2}
\end{equation}
The state $|B\rangle$ is a boundary state, which depends on the
particular model and boundary condition under consideration. In
the context of integrable models, it was introduced in the seminal
work of Ghoshal and Zamolodchikov \cite{Ghoshal:1993tm}. Here, we
take it with the normalisation $\bra 0|B\ket=1$. The factors
${\cal Z}_n$ and $\varep$ are defined through similar conditions
as in the bulk case.

In terms of the variables $r=|x_1-x_2|,\varep,m$, and using the
CFT normalisation of the branch-point twist  fields as well as
large-distance factorisation of correlation functions, we can
re-write the logarithmic divergence formula (\ref{divSA}) in a
more precise fashion. The IR (large-$mr$) and UV (small-$mr$)
leading behaviours of the entropy in the bulk are
 \beq\label{shlabu}
    S_A^{\rm bulk}(r) = \lt\{ \ba{ll} \frc{c}3 \log(r/\varepsilon) + o(1) & \varepsilon \ll r \ll m^{-1} \z
    -\frc{c}3 \log(\varepsilon m) + U^{\text{model}} + O((rm)^{-\infty})& \varepsilon \ll m^{-1} \ll r \ea\rt.
\eeq The term $U^{\text{model}}$ is a model-dependent
constant, which we computed exactly for the Ising model
\begin{equation}\label{u}
   U^{\text{Ising}}=-0.131984...
\end{equation}
in \cite{entropy} using QFT methods, reproducing results of
\cite{Jin}  obtained on the lattice. In (\ref{shlabu}), the
short-distance cut-off $\varep$ is of course non-universal, and it
is in general hard to evaluate its exact relation to the
correlation length. But the UV leading behaviour provides an
unambiguous QFT definition for it, independently of what the
correlation length of a particular quantum chain may be. Once this
definition is taken, the terms that are added to $-\frc{c}3
\log(\varepsilon)$ are universal -- these terms constitute the
universal part of the entanglement entropy. For instance, the
constant $U^{\text{model}}$ is indeed a universal QFT quantity.
Note that the IR leading behaviour is interpreted as the
saturation of the entanglement entropy at large distances.

In the boundary case, our definitions of the boundary state
further imply that the UV and IR leading behaviours of the
entanglement entropy are \beq\label{shla}
    S_A^{\rm boundary}(r) = \lt\{ \ba{ll} \frc{c}6 \log(2r/\varepsilon) + V(\kappa) + o(1) & \varepsilon \ll r \ll m^{-1} \z
    -\frc{c}6 \log(\varepsilon m) + \frc{U^{\text{model}}}2 + O((rm)^{-\infty})& \varepsilon \ll m^{-1} \ll r \ea\rt.
\eeq Again, $\varep$ is non-universal and hard to calculate. But
since we know that the constant $U^{\text{model}}$ is universal,
the IR behaviour in (\ref{shla}) gives an unambiguous definition
of $\varep$. Then, the terms that are added to $-\frc{c}6
\log(\varep)$ are the universal part of the entanglement entropy,
true QFT quantities. In particular, the UV behaviour in
(\ref{shla}) provides a universal definition for $V(\kappa)$. Note
that the leading asymptotic term at large distance
$U^{\text{model}}/2$ in the boundary case is just a choice. Once
this choice is made, the constant $V(\kappa)$ is universally
fixed. In particular, our short-distance cut-off $\varep$ here is
in general different in the bulk and boundary cases -- it is
related to the correlation length $\xi$ in different ways. Here,
$\kappa$ is a parameter that represents the boundary condition.

A major focus of our work \cite{entropy,other,nexttonext} has been
the study of the ratios of partition functions (\ref{result}) and
(\ref{result2}) at large distances $r=|x_1-x_2|$ (the infrared
(IR) region) for 1+1-dimensional integrable QFTs. Integrability
means that in these models there is no particle production in any
scattering process and that the scattering ($S$) matrix factorizes
into products of two-particle $S$-matrices which can be calculated
exactly (for reviews see e.g.
\cite{Karowski:1978eg,ZZ,abdalla,Mussardo:1992uc,Dorey:1996gd}).
This is the factorised-scattering theory of integrable models.
Since the scattering matrix and the particle spectrum fully
encodes the local definition of QFT, it is also possible to
incorporate the presence of boundaries in an integrable model
defined in the factorised-scattering way. The study of integrable
QFTs with boundaries has attracted a lot of attention in the last
two decades (see e.g.
\cite{Cherednik:1985vs,Sklyanin:1988yz,Fring:1993wt,Ghoshal:1993tm,Fring:1994ci,Bowcock:1995vp}).
In our work on the boundary case we have made extensive use of the
results of Ghoshal and Zamolodchikov \cite{Ghoshal:1993tm},
particularly the explicit realization of the boundary state which
they proposed.

Taking the known $S$-matrix of a model as input it is possible to
compute the matrix elements of local operators (also called form
factors). This is done by solving a set of consistency equations
\cite{KW,Smirnovbook}, also known as the form factor bootstrap
program for integrable QFTs. In \cite{entropy}, this program was
used and generalised in order to compute (\ref{result}) in the
case of integrable models with diagonal scattering matrix (that
is, without backscattering). In order to do this, the two-point
function in (\ref{result}) was expressed as a sum in terms of form
factors of the twist fields involved (an expansion using a decomposition in energy-momentum eigenstates). This was then extended to
models with backscattering, such as the sine-Gordon model
\cite{other}, to integrable models with boundaries in
\cite{nexttonext}, and some aspects were generalised to
non-integrable models \cite{next}.

One of the most interesting results of our works
\cite{entropy,other,next} has been the identification of the
next-to-leading order correction to the large-distance (large-$r$)
behaviour of the entropy of all (unitary) massive two-dimensional theories, that is,
the third term in the following large-$r$ expansion:
\beq\label{main}
    S_A^{\text{bulk}}(r) =-\frac{c}{3}\log(\varepsilon m)+U^{\text{model}}-  \frc18 \sum_{\alpha=1}^\ell K_0(2rm_\alpha) +
    O\lt(e^{-3rm}\rt).
\eeq Here, $m_\alpha$ are the masses of the $\ell$ particles in the QFT
model, with $m\equiv m_1\leq m_\alpha$, and $U^{\text{model}}$ is the
model-dependent constant introduced above. The first two terms are the expected
saturation of the entanglement entropy, but the interesting feature is the universal third
term, where we see that the leading exponential corrections are
independent of the scattering matrix, and only depend on the
particle spectrum of the model. This is quite striking: for
instance, a model of $\ell$ free particles of masses $m_\alpha$
will give the same leading exponential corrections as one with
interacting particles of the same masses. The result (\ref{main}) was first
obtained using integrable QFT methods \cite{entropy,other}, then, even more
strikingly, it was understood to hold as well outside of integrabilty \cite{next}.

Infrared corrections were
also studied for integrable models with boundary
\cite{nexttonext}, in which case they are always model-dependent, in particular
through the reflection matrices off the boundary.

The results
described so far have been extended further for the particular
case of the Ising model. For this model the particular form of all
infrared corrections to the entropy with and without boundary has
been obtained in \cite{nexttonext}. In fact, for this model without boundary,
like for other free-field QFT models,
there is an alternative powerful way of studying the entanglement entropy,
see the review \cite{CasiniHuertaRev}.

This paper is organized as follows. In section 2 we review the
relationship between partition functions on multi-sheeted Riemann
surfaces, correlation functions of branch-point twist fields and
the entanglement entropy. Employing
these relationships, we provide expressions for the bi-partite
entanglement entropy of 1+1-dimensional quantum field
theories, both in the bulk and boundary cases. In section 3 we
introduce the form factor program for branch-point twist fields.
We explain how it can be employed to obtain the form factors for
these twist fields in integrable models, and how these form factors can
be checked for consistency against conformal field theory results.
For the two-particle form factors we generalize this program also
to 1+1-dimensional non-integrable QFTs. We identify the general
structure of the two particle form factors
and, in integrable cases, suggest how higher particle form factors may be obtained from
lower particle ones. For the Ising model, we give closed formulae
for all higher particle form factors. In general, we recall how
the form factors can be regarded as
building blocks for correlation functions.
The correlation functions can then be expressed as series where
the leading contributions at large distances arise from the lower
particle form factors. In section 4, we used these form factor series
in order to analyse the entanglement entropy, both in the bulk and boundary
cases. In the bulk case, our most important result
is the universal expression (\ref{main}) for the next-to-leading order correction
to the entanglement entropy at long distances.
Both for the bulk and boundary cases of the Ising model, we
evaluate all higher order infrared corrections to the entanglement
entropy. In the boundary case, we find the precise relationship
between the ultraviolet leading behaviour of the entanglement entropy and the
boundary entropy introduced by Affleck and Ludwig
\cite{Affleck:1991tk}. Finally, in section 5 we summarise our
conclusions and outlook.

\sect{Replica trick and entanglement entropy}

\subsection{Partition functions on multi-sheeted Riemann surfaces}
\label{sspartfunct}

From the considerations
in the introduction, it is clear that for the study of the
entanglement entropy in the scaling limit, we must study partition
functions of (euclidean-signature) quantum field theory on
multi-sheeted Riemann surfaces. We will come back to the precise
relation in the next section, but for now, let us discuss such
partition functions. In particular, we wish to introduce the
concept of branch-point twist fields, following \cite{entropy}.

The partition function of a model of 1+1-dimensional QFT, with
local lagrangian density ${\cal L}[\varphi](x)$, on a Riemann
surface ${\cal R}$ is formally obtained by the path integral \beq
\label{partfunct}
    Z[{\cal L},{\cal R}] = \int [d\varphi]_{{\cal R}} \exp\lt[-\int_{{\cal R}} d^2x_{{\cal R}}\,{\cal L}[\varphi](x_{{\cal R}})\rt].
\eeq Here, $[d\varphi]_{{\cal R}}$ is an infinite measure on the
set of configurations of some field $\varphi$ living on the
Riemann surface ${\cal R}$ and on which the lagrangian density
depends in a local way, and $x_{{\cal R}}$ is a point on the
Riemann surface. Consider Riemann surfaces with zero curvature
everywhere except at a finite number of points; the points where the curvature is non-zero are branch points.
In the case where
the initial quantum system we consider is on the line, the Riemann
surface forms a multiple covering of $\R^2$; in the case where the
model is on the half-line, it forms a multiple covering of the
half-plane, which we will take to be the right half-plane,
$\R^2_\rightarrow \equiv \{(\rx,\ry)\,|\,\rx>0,\,\ry\in\R\}$. For
simplicity, we will only consider the case $\R^2$ in the following
discussion, but the case $\R^2_\rightarrow$ is entirely similar.
Since the lagrangian density does not depend explicitly on the
Riemann surface as a consequence of its locality (and the fact
that the curvature is zero almost everywhere), it is expected that
this partition function can be expressed as an object calculated
from a model on $\R^2$, where the structure of the Riemann surface
is implemented through appropriate boundary conditions around the
points with non-zero curvature. Consider for instance the simple
Riemann surface $\orb_{n,x_1,x_2}$ with $x_j = (\rx_j,0)$,
composed of $n$ sheets sequentially joined to each other on the
segment $\rx\in[\rx_1,\rx_2],\,\ry=0$ (see Fig. \ref{fig-sheets}). We would expect that the associated
partition function involves certain ``fields''\footnote{Here, the
term ``field'' is taken in its most general QFT sense: it is an
object of which correlation functions -- multi-linear maps -- can
be evaluated, and which depends on a position in space --
parameters $\rx,\ry$ that transform like coordinates under
translation symmetries.} at the points on $\R^2$ given by $x_1$
and $x_2$. These fields would be implementing the appropriate
non-zero curvature at the branch points.

The expression (\ref{partfunct}) for the partition function
essentially defines these fields: it gives their correlation
functions, up to a normalisation independent of their positions;
we only have to put as many branch points as we have such fields
in the correlation function. But if we insist in understanding
this as the initial model on $\R^2$, this definition makes these
special fields non-local. Locality of a field (used here in its
most fundamental sense) means that as an observable in the quantum
theory, it is quantum mechanically independent of the energy
density at space-like distances. In the associated euclidean field
theory, this means that correlation functions involving these
fields and the energy density are, as functions of the position of
the energy density, defined on $\R^2$ and continuous except at the
positions of the fields. The energy density is simply obtained
from the lagrangian density. But clearly, bringing the lagrangian
density around  a branch point changes the value of the
correlation function, since it gets to a different Riemann sheet.
Hence, the fields defined by (\ref{partfunct}) seen as fields in a
model on $\R^2$ with lagrangian density ${\cal L}[\varphi](x)$ are
non-local: the lagrangian density is not well-defined on $\R^2$.
Locality is at the basis of most of the results in QFT, so it is
important to recover it.

In order to correct the problem, note that if we {\em defined} a new lagrangian density on $\R^2$ at the point $(\rx,\ry)$ by simply summing the initial lagrangian density $\sum_{j=1}^n {\cal L}[\varphi](x^{(j)}_{{\cal R}})$ over all the points $x^{(j)}_{{\cal R}}$ of the Riemann surface ${\cal R} = \orb_{n,x_1,x_2}$ that project onto $(\rx,\ry)$, then with respect to this new lagrangian density, the fields defined above would be local.
Hence, the idea is simply to consider a larger model: a model formed by
$n$ independent copies of the original model, where $n$ is the
number of Riemann sheets necessary to describe the Riemann surface
by coordinates on $\R^2$. Let us take again the simple example of
$\orb_{n,x_1,x_2}$. We re-write (\ref{partfunct}) as
\beq\label{partfunctmulti}
    Z_n(x_1,x_2)\equiv Z[{\cal L},\orb_{n,x_1,x_2}] =
        \int_{{\cal C}(\rx_1;\rx_2)} [d\varphi_1 \cdots d\varphi_n]_{\R^2}\, e^{-\int_{\R^2} d^2x\,(
        {\cal L}[\varphi_1](x)+\ldots+{\cal L}[\varphi_n](x))},
\eeq where ${\cal C}(\rx_1,\rx_2)$ are {\em continuity conditions} on the fields
$\varphi_1,\ldots,\varphi_n$ restricting the path integral: \beq
    {\cal C}(\rx_1;\rx_2) \quad:\quad \varphi_i(\rx,0^+) = \varphi_{i+1}(\rx,0^-)~,\quad \rx\in[\rx_1,\rx_2],\;
    i=1,\ldots,n, \label{this}
\eeq where we identify $n+i\equiv i$. What appears in the action
of that path integral can now be seen as the lagrangian density of
the multi-copy model,
\[
    {\cal L}^{(n)}[\varphi_1,\ldots,\varphi_n](\rx,\ry) = {\cal L}[\varphi_1](\rx,\ry)+\ldots+{\cal L}[\varphi_n](\rx,\ry).
\]
The energy density of that model is the sum of the energy
densities of the $n$ individual copies. The expression
(\ref{partfunctmulti}) with (\ref{this}) does indeed define the insertion of local fields at
$x_1$ and $x_2$ in the multi-copy model, since the energy density is
the same on both sides of the segment $\rx\in[\rx_1,\rx_2],\,\ry=0$
according to the conditions ${\cal C}(\rx_1,\rx_2)$.

The local fields defined by (\ref{this}) are examples of ``twist
fields''. Twist fields exist in a QFT model whenever there is a
global internal symmetry $\sym$ (a symmetry that acts the same way
everywhere in space, and that does not change the positions of
fields): $\int_{\R^2}d\rx d\ry\, \t{\cal L}[\sym\varphi](\rx,\ry)
= \int_{\R^2}d\rx d\ry\, \t{\cal L}[\varphi](\rx,\ry)$ for some
lagrangian density $\t{\cal L}[\varphi](x)$. Their correlation
functions can be formally defined through the path integral:
\beq\label{formdef}
    \bra \tw_\sym(x) \cdots \ket_{\t{\cal L},\R^2} \propto
        \int_{{\cal C}_\sym(x)} [d\varphi]_{\R^2} \exp\lt[-\int_{\R^2} d\rx d\ry\,\t{\cal L}[\varphi](\rx,\ry)\rt] \cdots
\eeq where $\cdots$ represent insertions of other local fields at
different positions. The path integral continuity conditions produce a cut on a half-line starting at the point $x=(\rx,\ry)$: \beq
    {\cal C}_\sym(x) \quad:\quad \varphi(\rx',\ry^+) = \sym\varphi(\rx',\ry^-)~,\quad \rx'\in[\rx,\infty)~.
\eeq The proportionality constant is an infinite constant that is
independent of the position $x$ and of those of the other local
fields inserted, present in order to render the path integral
finite (so that it may represent a correlation function). For
insertion of many twist fields, we just add more continuity
conditions on different half-lines, starting at different points. The fact that $\sym$ is a
symmetry ensures that $\tw_\sym$ is local, since it ensures that
the energy density is continuous through the cut produced by the
continuity condition. Also, it ensures that the result is in fact
invariant under continuous changes of the shape of this cut, up to
symmetry transformations $\sym$ of the local fields that are being
swept. This is because inside a loop, we can always apply a
symmetry transformation without changing the result of the path
integral, up to transformations of local fields present inside the
loop. In this way, we can modify the continuity conditions through
the loop. By drawing a line, where the usual continuity holds,
starting and ending on a twist-field cut (or possibly on its
end-points), we form a loop with the twist-field cut. Applying a
symmetry transformation inside this loop, we erase part (or all)
of the twist-field cut and make the line a new twist-field cut. In
this way, it is possible to move the twist-field cut to any shape.
Hence our choice of a half-line extending to the right in the
definition of the twist field is just for convenience.

A consequence of the formal definition (\ref{formdef}) is that
correlation functions $\bra \tw_\sym(x) \Or(y) \cdots\ket_{\t{\cal
L},\R^2}$ with some local fields $\Or(y)$ are defined, as
functions of $y$ (continuous except at positions of other local
fields), on a multi-sheeted covering of $\R^2$ with a branch point
at $y=x$, whenever $\sym\Or \neq\Or$. More precisely, twist fields
have the property that a clockwise continuous displacement of a
local field $\Or(y)$ around  $x$ back to its initial projected
point on $\R^2$ is equivalent to the replacement $\Or\mapsto
\sym\Or$ in any correlation function. If $\sym\Or \neq\Or$, then
$\Or$ is said to be ``semi-local'' with respect to $\tw_\sym$.
This twist-field property is satisfied by a large family of
fields, not only the one obtained through the formal definition
(\ref{formdef}). For instance, we could have inserted a field
$\varphi(x)$ in the path integral, leading to the same twist
property (this is a descendent twist field). However, the twist
property, along with the condition that $\tw_\sym$ has the lowest
scaling dimension and be invariant under all symmetries of the
model that commute with $\sym$ (that is, that it be a primary
field in the language of conformal field theory), uniquely fixes
the field $\tw_\sym$ up to a normalisation. These conditions lead
to a definition that is in agreement with the path-integral
definition (\ref{formdef}). In fact, these conditions constitute a more fundamental
way of defining the primary twist field than the path integral, as
they do not require the existence of a lagrangian density. In
particular, they lead to unambiguous definitions in any
quantisation scheme (we will discuss what the twist condition is
in the quantisation on the line when we discuss form factors). We
will take this general point of view in the following, but we will
continue to think of a model of QFT through its lagrangian density
$\t{\cal L}$ for clarity.

Twist fields associated to internal symmetries have been largely studied in the
context of CFT: they correspond to twisted modules for
vertex operator algebras \cite{FLM,FFR,D}, at the basis of so-called orbifold models.
In the context of massive integrable
QFT, only the simplest (standard) cases are well known. The Ising order and disorder fields are $\Z_2$
twist fields in the free massive Majorana fermion theory, and in
the equivalence between the massive Thirring model and the sine-Gordon model,
the bosonic exponential fields of the latter are $U(1)$ twist fields of the former.
But in fact, for fields with more general twist propertes,
the form factor equations of integrable QFT were written in \cite{BernardL92}.
The first extensive study of form factors of non-standard twist fields (see below)
was done in \cite{entropy,other},
and the first time twist fields form factors were considered beyond integrability was in \cite{next};
we will describe these works in the next section.

\begin{figure}
\bc
\includegraphics[width=6cm,height=3cm]{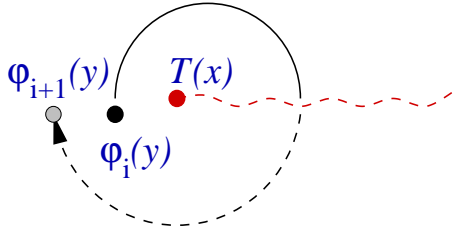}
\ec \caption{The effect of $\tw$ on other local fields.}
\label{fig-Cx}
\end{figure}

In the $n$-copy model with lagrangian $\t{\cal L} = {\cal L}^{(n)}$, there is a symmetry
under exchange of the copies. The twist fields defined by
(\ref{partfunctmulti}), which we call {\em branch-point twist
fields}, are twist fields associated to the two opposite cyclic
permutation symmetries $i\mapsto i+1$ and $i+1\mapsto i$
($i=1,\ldots,n,\;n+1\equiv 1$). We will denote them simply by
$\tw$ and $\t\tw$, respectively: \beqa &&
    \tw=\tw_\sym~,\quad \sym\;:\; i\mapsto i+1 \ {\rm mod} \,n \n &&
    \t\tw=\tw_{\sym^{-1}}~,\quad \sym^{-1}\;:\; i+1\mapsto i \ {\rm mod} \,n\no
\eeqa (see Fig. \ref{fig-Cx} for the case $\tw$).
More precisely, we have \beq\label{pftwopt}
    Z_n(x_1,x_2) \propto \bra \tw(x_1) \t\tw(x_2) \ket_{{\cal L}^{(n)},\R^2}.
\eeq This can be seen to be correct by observing that for
$\rx\in[\rx_1,\rx_2]$, consecutive copies are connected through
$\ry=0$ due to the presence of the cut produced by $\tw(x_1)$,
whereas for $\rx>\rx_2$, the additional cut produced by
$\t\tw(\rx_2,0)$ cancels this, and copies are connected to
themselves through $\ry=0$.  Hence, indeed only a finite cut
between $\rx_1$ and $\rx_2$ remains. The precise proportionality
constant was discussed around equation (\ref{result}).

More generally, the identification holds for correlation functions
in the model ${\cal L}$ on $\orb_{n,x_1,x_2}$, this time with an
equality sign: \beq\label{br-tw}
    \bra \Or(y_{{\cal R}} \mbox{\ on sheet $i$}) \cdots \ket_{{\cal L},\orb_{n,x_1,x_2}} =
    \frc{\bra \tw(x_1) \t\tw(x_2) \Or_i(y) \cdots \ket_{{\cal L}^{(n)},\R^2}}{\bra \tw(x_1) \t\tw(x_2) \ket_{{\cal L}^{(n)},\R^2}}
\eeq where $\Or_i$ is the field in the model ${\cal L}^{(n)}$
coming from the $i^{\rm th}$ copy of ${\cal L}$, and $y$ is the projection of $y_{{\cal R}}$ onto $\R^2$.

Note that it is easy to transform the twist field $\tw$ into
$\t\tw$, and vice versa: we only have  to apply the ``flip''
symmetry transformation by which the order of the copies is
inverted (this is another element of the permutation symmetry
group). Note also that this construction can also be generalised
to Riemann surfaces with more branch points is straightforward,
but this will not be needed here.

The conformal dimension of branch-point twist fields is an
important characteristic of these fields. It was essentially
calculated in \cite{Calabrese:2004eu}, although branch-point twist
fields were not introduced; only the non-local fields discussed
above were considered. Here, we reproduce the derivation  of
\cite{entropy}, which makes explicit reference to the branch-point
twist fields, but otherwise follows closely
\cite{Calabrese:2004eu}.

Consider the model ${\cal L}$ to be a conformal field theory
(CFT). Then also ${\cal L}^{(n)}$ is a CFT. There are $n$ fields
$T_j(w)$ in ${\cal L}^{(n)}$ that correspond to the holomorphic stress-energy
tensors of the $n$ copies of ${\cal L}$, and in particular the sum
$T^{(n)}(w) = \sum_{j=1}^n T_j(w)$ is the holomorphic stress-energy tensor of
${\cal L}^{(n)}$. The central charge of ${\cal L}^{(n)}$ is $nc$,
if $c$ is that of ${\cal L}$.

Consider the holomorphic stress-energy tensor $T(w)$ in ${\cal L}$. We can
evaluate the one-point function $\bra T(w)\ket_{{\cal
L},\orb_{n,x_1,x_2}}$ by making a conformal transformation from
$z$ in $\R^2$ to $w$ in $\orb_{n,a_1,a_2}$ (here $z$ and $w$ are
complex coordinates, and $w_j = \rx_j+i \ry_j$) given by
\[
    z = \lt(\frc{w-w_1}{w-w_2}\rt)^{\frc1n}~.
\]
We have
\[
    \bra T(w)\ket_{{\cal L},\orb_{n,x_1,x_2}} = \lt(\frc{\p z}{\p w} \rt)^2 \bra T(z)\ket_{{\cal L},\R^2} + \frc{c}{12} \{z,w\}
\]
where the Schwarzian derivative is
\[
    \{z,w\} = \frc{z''' z' - (3/2) (z'')^2}{(z')^2}~.
\]
Using $\bra T(z)\ket_{{\cal L},\R^2}=0$, we obtain
\[
    \bra T(w)\ket_{{\cal L},\orb_{n,x_1,x_2}} = \frc{c(n^2-1)}{24 n^2} \frc{(w_1-w_2)^2}{(w-w_1)^2(w-w_2)^2} ~.
\]
Since, by (\ref{br-tw}), this is equal to $\bra \tw(x_1)
\t\tw(x_2) T_j(w)\ket_{{\cal L}^{(n)},\R^2}/\bra \tw(x_1)
\t\tw(x_2)\ket_{{\cal L}^{(n)},\R^2}$ for all $j$, we can
evaluate the correlation function involving the stress-energy
tensor of ${\cal L}^{(n)}$ by multiplying by $n$:
\[
    \frc{\bra \tw(x_1) \t\tw(x_2) T^{(n)}(w)
    \ket_{{\cal L}^{(n)},\R^2}}{\bra \tw(x_1) \t\tw(x_2)\ket_{{\cal L}^{(n)},\R^2}} =
    \frc{c(n^2-1)}{24 n} \frc{(w_1-w_2)^2}{(w-w_1)^2(w-w_2)^2}~.
\]
From the usual CFT formula for insertion of a stress-energy tensor
\beqa &&
    \bra \tw(x_1) \t\tw(x_2) T^{(n)}(w)\ket_{{\cal L}^{(n)},\R^2} =
    \n && \qquad
    \lt(\frc1{w-w_1} \frc{\p}{\p w_1} + \frc{h_1}{(w-w_1)^2} + \frc1{w-w_2} \frc{\p}{\p w_2} + \frc{h_2}{(w-w_2)^2}\rt)
    \bra \tw(x_1) \t\tw(x_2)\ket_{{\cal L}^{(n)},\R^2} \no
\eeqa we identify the scaling dimension of the primary fields
$\tw$ and $\t\tw$ (they have the same scaling dimension) using
$\bra \tw(x_1) \t\tw(x_2)\ket_{{\cal L}^{(n)},\R^2} =
|w_1-w_2|^{-2d_n}$: \beq\label{scdim}
    d_n = \frc{c}{12} \lt(n-\frc1n\rt)~.
\eeq This dimension is the lowest possible
dimension for fields with the branch-point twist property, as it is the dimension of a primary field in this family.

Hence, the branch-point twist fields $\tw$ and $\t\tw$ are
unambiguously defined by specifying their associated symmetry,
$\sym$ and $\sym^{-1}$, namely that they are invariant under other
global symmetry transformations that commute with $\sym$, and by
specifying their scaling dimension to be $d_n$ (\ref{scdim}). This
is a definition up to normalisation; we will come back to the
normalisation in the next section. There are of course many other
twist fields associated to other elements of the permutation
symmetry group, and these may have smaller scaling dimension than
$d_n$, but these are not the branch-point twist fields that are
used in order to represent the partition function on multi-sheeted
Riemann surfaces.

\subsection{Bulk entanglement entropy}

Partition functions on Riemann surfaces with branch
points can be used in order to evaluate the entanglement entropy.
We start by considering the bulk case, where the quantum model is
on the line, and the corresponding (euclidean) QFT model is on
$\R^2$. We are ultimately interested in the scaling limit of the
ground-state entanglement entropy, (\ref{defeegs}). However, we
will provide here general arguments valid not only for the ground
state, by also for excited states. Hence, in place of $|{\rm
gs}\ket$ in (\ref{defeegs}), we will take some arbitrary, possibly
excited state $|\psi\ket$. In the scaling limit, the ground state
$|{\rm gs}\ket$ maps to the QFT vacuum state $|0\ket$, but excited
states map to QFT asymptotic states characterised by the particle
content and their momenta (in general, we will keep the notation
$|\psi\ket$ for the corresponding state in QFT). We will later
specialise $|\psi\ket$ to the ground state.

The main idea in order to evaluate the entanglement entropy in the
scaling limit of a quantum model is to use the replica trick
\cite{CallanW94,HolzheyLW94,Calabrese:2004eu,Calabrese:2005in}; we
will  follow more precisely the method of \cite{Calabrese:2004eu}.
That is, we use the identity (\ref{formulan1}), where we start by
considering the trace for positive integer $n$, evaluate it in QFT
(in the scaling limit), then take the appropriate analytic
continuation in $n$ in order to evaluate the limit of the
derivative. Considering positive integer $n$ in QFT is useful,
because there, the trace is directly related to a partition
function on a multi-sheeted Riemann surface, studied in the
previous subsection. The way this works is as follows.

First, in order to construct $\rho_A$ in the scaling limit,
consider  the QFT Hilbert space as a space of field configurations
$\{\varphi(\rx),\,\rx\in\R\}$ on $\R$. A state $|\psi\ket$ can be
written as a linear combination of field configurations on $\R$
where the coefficients are obtained by path integrals on the lower
half of the $\R^2$ plane, $\R^2_\downarrow =
\{(\rx,\ry)\,|\,\rx\in\R,\,\ry<0\}$. The boundary condition on
$\ry=0$ is determined by the coefficient we are looking at, and
the asymptotic condition $\ry\to0$ is determined by $|\psi\ket$.
That is, \beq\label{psiexp}
    |\psi\ket =\int [d\phi]_\R\; |\phi\ket\bra\phi|\psi\ket ,\quad
    \bra \phi|\psi\ket =  \frc1{\sqrt{Z_1}}\int_{{\cal C}} [d\varphi]_{\R^2_\downarrow} \;e^{-S_{\R^2_\downarrow}[\varphi]}
\eeq
where $S_{\cal M}[\varphi] = \int_{{\cal M}} d^2x\,{\cal L}[\varphi](x)$ is the action of the model on ${\cal M}$, and the condition ${\cal C}$ is
\[
    {\cal C}\ :\  \lt\{\varphi(\rx,0)=\phi(\rx),\,\rx\in\R\atop \varphi(\rx,\ry\to-\infty)\sim
    f_\psi(\rx,\ry),\,\rx\in \R\rt\}.
\]
Here, $f_\psi(\rx,\ry)$ represents the asymptotic condition
corresponding  to the state $|\psi\ket$; for instance,
$f_{\psi}(\rx,\ry)=0$ for the vacuum state, otherwise it
reproduces wave packets corresponding to asymptotic particles. The
number $Z_1$ is the ``partition function'' of the QFT model in the
state $|\psi\ket$:
\[
    Z_1 = \int_{{\cal C}'} [d\varphi]_{\R^2} \;e^{-S_{\R^2}[\varphi]}
\]
where
\[
    {\cal C'}\ :\  \lt\{\varphi(\rx,\ry\to-\infty)\sim
    f_\psi(\rx,\ry),\,\rx\in \R\atop \varphi(\rx,\ry\to\infty)\sim
    f_\psi^*(\rx,\ry),\,\rx\in \R\rt\}.
\]
It is a true partition function if $|\psi\ket$ is the ground
state; otherwise,  it corresponds to an excited-state
normalisation. The factor $1/\sqrt{Z_1}$ is inserted here in order
that we maintain the proper normalisation $\bra\psi|\psi\ket=1$
(the states $|\phi\ket$ are taken with the natural
delta-functional normalisation). Note that the coefficients in the
expansion of the dual vector $\bra\psi|$ are simply obtained from
(\ref{psiexp}) by complex conjugation, along with the inversion of
the imaginary time $\ry\mapsto-\ry$ -- this naturally gives a path
integral on the other half of $\R^2$.

Next, the density matrix $\rho_A$ is constructed by tracing
$|\psi\ket\bra\psi|$ over degrees  of freedom on $\b{A}$. We sum
the diagonal matrix elements of $|\psi\ket\bra\psi|$, but only in
the space $\b{\cal A}$, formed by field configurations on $\b{A}$.
This has the effect of ``connecting'', on the part $\b{A}$ of
their boundaries, the two surfaces on which the path integrals
used to represent matrix elements of $|\psi\ket$ and of
$\bra\psi|$ are defined. Hence, we get a path integral on the
whole $\R^2$, with continuity on $\b{A}$, but with an open slit on
$A$. We are left with a matrix element of $\rho_A$, determined by
field configurations on both sides of the slit on $A$. That is, we
find \beq
    \bra \phi_A|\rho_A|\phi_A'\ket =
    \int [d\phi_{\b{A}}]_{\b{A}}\; \bra \phi_{\b{A}},\phi_A|\psi\ket\bra\psi|\phi_{\b{A}},\phi_A'\ket
    = \frc1{Z_1}\int_{{\cal C}''} [d\varphi]_{\R^2\setminus(A,0)}\; e^{-S_{\R^2\setminus (A,0)}[\varphi]}
\eeq
where
\[
    {\cal C}'' = \lt\{
    \varphi(\rx,0^-)=\phi_A(\rx),\,\rx\in A;\ \varphi(\rx,0^+)=\phi_A'(\rx),\,\rx\in A\atop
    \varphi(\rx,\ry\to-\infty)\sim
    f_\psi(\rx,\ry),\,\rx\in\R;\ \varphi(\rx,\ry\to\infty)\sim
    f_\psi^*(\rx,\ry),\,\rx\in\R\rt\}.
\]
The $n^{\rm th}$ power of $\rho_A$,
\[
    \bra \phi_A|\rho_A^n|\phi_A'\ket = \int [d\phi_1 d\phi_2 \cdots d\phi_{n-1}]_A
    \bra \phi_A|\rho_A|\phi_1\ket
        \bra \phi_1|\rho_A|\phi_2\ket \cdots \bra \phi_{n-1}|\rho_A|\phi_A'\ket,
\]
can be obtained by taking $n$ copies of $\rho_A$ with their own independent path integrals over $\varphi_j,\,j=1,\ldots,n$, and by connecting in a sequential way one side of the slit on $A$ of one copy to the other side of the slit on $A$ of the next copy (that is, having continuity of the fields $\varphi_j$ through the slit $A$ in this sequential way). Finally, taking the trace connects the last copy to the first, so that we obtain the partition function on a multi-sheeted Riemann surface. Taking $A$ to consist of only one interval for simplicity, with end-points at positions $x_1$ and $x_2$ in $\R^2$, we then find, in agreement with (\ref{iden}), \beq \label{TrAZn}
    \Tr_{{\cal A}} \rho_A^n = \frc{Z_n(x_1,x_2)}{Z_1^n}
\eeq where the partition function is (\ref{partfunctmulti}). In
this derivation, we referred to the definition of the QFT model
given by its lagrangian density ${\cal L}$, but as mentioned in
the previous subsection, this is just for clarity of exposition;
the existence or not of a lagrangian density does not affect any
of the results.

As we saw in the previous subsection, the partition function above can be computed as a two-point correlation function of
local fields in the model given by the lagrangian density ${\cal L}^{(n)}$ using (\ref{pftwopt}). In the case of the ground state, using (\ref{result}), we have
\beq\label{rhon}
    S_A^{\text{bulk}}(r) =
    -\lim_{n\rightarrow 1}\frac{d}{dn}{\cal Z}_n\varep^{2d_n} \bra0| \tw(x_1) \t\tw(x_2) |0\ket.
\eeq In the case where $|\psi\ket$ is an excited state, the
relation (\ref{pftwopt}) still holds, because the derivation of
the previous subsection did not make any reference to the
asymptotic conditions on the field $\t\varphi$ as
$\ry\to\pm\infty$. However, in (\ref{pftwopt}), we have to
understand the correlation function as an excited-state diagonal
matrix element of the product of twist fields, in order to
implement the appropriate asymptotic conditions in the path
integral. Hence, what replaces (\ref{result}) is
\beq\label{resultexc}
    \frc{Z_n(x_1,x_2)}{Z_1^n} = {\cal Z}_n\varep^{2d_n} \bra\psi| \tw(x_1) \t\tw(x_2) |\psi\ket.
\eeq
so that we obtain
\beq\label{rhonexc}
    S_A^{\text{bulk}}(r) =
    -\lim_{n\rightarrow 1}\frac{d}{dn}{\cal Z}_n\varep^{2d_n} \bra\psi| \tw(x_1) \t\tw(x_2) |\psi\ket.
\eeq In this formula, the normalisation of $|\psi\ket$ is not the
standard one in the case of  an excited state, because of the
normalisation factor $1/Z_1^n$ in (\ref{resultexc}). Indeed, this
factor normalises away the infinite-volume divergencies coming
from colliding rapidities of asymptotic states in the disconnected
terms of the two-point function. We will not discuss further here
the case of excited states, and concentrate solely on the ground
state entanglement entropy.

\subsection{From bulk to boundary entanglement entropy}\label{bent}

We are now interested in the situation where the system
is on the half-line $\rx>0$ and the connected region $A$ has an
end-point on the boundary of the system $\rx=0$. However, it will
be most easy and instructive to start with a slightly different
problem, where the region $A$ lies entirely in the bulk, with two
boundary points on the half line $\rx>0$. Since the system is on
the half-line, one needs to provide an additional boundary
condition at $\rx=0$ in order to fully define the model. There are
various ways of implementing such a boundary condition. From the
viewpoint of the path integral, the boundary condition is
implemented by a restriction on the allowed values of the fields
$\varphi(x)$ and its derivatives at the boundary $\rx=0$, along
with, possibly, an extra term $S_B$ in the action that is
supported on $\rx=0$, $S_B = \int d\ry {\cal L}_B[\varphi](\ry)$.
From the viewpoint of the quantised theory on the half-line
$\rx>0$, with $\ry\in\R$ the imaginary time, the boundary
condition determines the whole Hilbert space, the vacuum
$|0\ket_B$ and all excited states. Finally, crossing symmetry
gives, from the latter, the viewpoint of the quantised theory on
the full line $\ry\in\R$ with imaginary time $\rx>0$. There, the
boundary condition corresponds to a boundary state $|B\ket$, a
state in the usual Hilbert space on the full line.

The derivations of the previous two subsections, connecting the
entanglement  entropy to partition functions on multi-sheeted
Riemann surfaces, and then connecting the latter to correlation
functions of branch-point twist fields, can be directly
generalised to the boundary situation. First, retracing the steps
of the last subsection in the path-integral formulation of the
model on the boundary, we obtain again (\ref{TrAZn}), where now
the partition functions are path integrals over configurations of
the field $\t\varphi(x)$ with $x\in\R_\rightarrow^2$, with a
boundary condition at $\rx=0$ and possibly an extra boundary term
$S_B$ in the action. Second, the derivation of subsection
\ref{sspartfunct} can also be directly reproduced, and we obtain,
instead of (\ref{result}), \beq\label{resultbound}
    \frc{Z_n(x_1,x_2)}{Z_1^n} = {\cal Z}_n\varep^{2d_n} {}_B\bra0| \tw(x_1) \t\tw(x_2) |0\ket_B.
\eeq Here $|0\ket_B$ is the ground state in the $n$-copy QFT model
on the half line, and $\tw$ and $\t\tw$ are the twist fields with
the same fundamental definition as before (the twist property,
invariance under other symmetry transformations, and lowest
scaling dimension), but as operators on the half-line Hilbert
space. Hence we obtain \beq\label{rhonbound}
    S_A^{\text{boundary}}(x_1,x_2) =
    -\lim_{n\rightarrow 1}\frac{d}{dn}{\cal Z}_n\varep^{2d_n} {}_B\bra0| \tw(x_1) \t\tw(x_2) |0\ket_B.
\eeq

From this, there are two ways to obtain the entanglement entropy
$S_A^{\text{boundary}}(r)$ for a region $A$ starting from $\rx=0$
and ending at a distance $r$ from it. First, we could consider the
limit $x_1\to(0,0)$ in $S^{\text{boundary}}_A(x_1,x)$ with
$x=(r,0)$, making the region $A$ approach the boundary. As the
twist field at $x_1=(\rx_1,0)$ approaches the boundary, the
correlation function diverges, because the presence of the
boundary changes the regularisation necessary around the branch
point. A way to evaluate the divergency is to use boundary
conformal field theory, which applies in massive models when a
local field is near to a boundary. It tells us that for small
$\rx_1$ there is a power law determined by the conformal dimension
of ${\cal T}$: ${}_B\bra0| \cdots {\cal T}(x_1)|0\ket_B \propto
\rx_1^{-d_n}$ \cite{CardyLewellen}. We may then define ${\cal
T}(0)|0\ket_B$ as $\lim_{\rx_1\to0} \rx_1^{d_n} {\cal
T}(x_1)|0\ket_B$. This appropriately regularised operator ${\cal
T}(0)$ is simply proportional to the unitary operator performing a
cyclic $\Z_n$ transformation, since its branch cut, through which
$\Z_n$ transformations are performed, now extends through the
whole space. But since $|0\ket_B$ is invariant under such a
transformation, the action of ${\cal T}(0)$ on $|0\ket_B$ gives
$|0\ket_B$ back.  Moreover, since in the definition of $\tw(0)$ we
already took into account a renormalisation of the field, there is
no regularisation $\varep^{d_n}$ associated to that field in the
relation between the partition function and the two-point
function. Hence, we find, with appropriate choice of
proportionality constants,
\begin{equation}\label{be2}
   S_A^{\rm boundary}(r)=
   - \lim_{n \rightarrow 1}\frac{d}{d n} {\cal Z}_n \varepsilon^{d_n}{}_B\langle 0|\tw(x)|0\rangle_B.
\end{equation}
The power of $\varep$ guarantees that the scaling dimension of the quantity that is differentiated is 0.

It is important to note that the result of the limit $x_1\to(0,0)$
is a function  only of the distance $r$ between $x$ and the
boundary. Indeed, ${\cal T}(0)$ is a unitary operator preserving
$|0\ket_B$, hence can be put at any imaginary time without
changing the result.

Second, we may take the limit $\rx_2\to\infty$ in
$S^{\text{bulk}}_A(x,x_2)$, with again $x=(r,0)$. This should lead
to the same  result for the entanglement entropy, because of the
symmetry $S_A = S_{\b{A}}$, up to an additive term corresponding
to the contribution to the entanglement entropy around the
boundary point at $\infty$. In the limit $\rx_2\to\infty$, the
two-point function in (\ref{rhon}) reduces to its disconnected
part: \beq
    {}_B\langle 0|\tw(x)\t\tw(x_2)|0\rangle_B \sim {}_B\langle 0|\tw(x)|0\ket_B {}_B\bra0| \t\tw(\infty)|0\rangle_B~.
\eeq In the second factor, the twist field does not feel the
presence of the boundary, hence this expectation value can be
replaced by its expectation value in the model without boundary,
$\bra 0| {\cal T}|0 \ket$. Dividing out this factor (which
corresponds to subtracting the contribution to the entanglement
entropy around the point at $\infty$), flipping the sheets in
order to transform $\t\tw$ into $\tw$, and using the appropriate
branch-point regularisation, we find again (\ref{be2}).

Some of these considerations, and in particular the calculations
that we will present below, are made clearer by using crossing
symmetry: taking the half-line $\rx>0$ to be imaginary time, and
the full line $\ry\in\R$ the space direction. In this picture, as
we mentioned, the boundary condition is implemented as a boundary
state on the Hilbert space on the full line: \beq
    {}_B\langle 0|{{{\cal T}}}(x_1) \tilde{{\cal
    T}}(x_2)|0\rangle_B =  \langle 0|{{{\cal T}}}(x_1) {\tilde{\cal T}}(x_2)|B\rangle.
\eeq The boundary state $|B\ket$ is in the past at imaginary time
$\rx=0$ in the Hilbert space of the model on the full line, and
the twist fields are placed at imaginary times $\rx_1$ and
$\rx_2$. More precisely, the boundary state is the $n$-fold tensor
product of single-copy boundary states. The state $\bra0|$ is
again the ground state of the $n$-copy model on the line,
corresponding to asymptotic vanishing fields conditions at
positive infinite times. No factor occurs in using crossing
symmetry since the branch-point twist fields are spinless. The
normalisation of the boundary state $|B\ket$ is such that
$\bra0|B\ket = 1$. By any of the two ways explained above, we
obtain for the entanglement entropy:
\begin{equation}\label{be3}
   S_A^{\rm boundary}(r)=
   - \lim_{n \rightarrow 1}\frac{d}{d n} {\cal Z}_n \varepsilon^{d_n}\langle 0| {{\cal T}}(x)|B\rangle.
\end{equation}
where again the distance between the point $x$ and the boundary is $r$.

\subsection{Normalisations of branch-point twist fields and boundary
states}\label{ssectnorm}

In order to obtain a universal result for the
entanglement entropy, we need to fix the short-distance behaviour
as in (\ref{shlabu}) in the bulk case, and fix the large-distance
behaviour as in (\ref{shla}) in the boundary case. As was
explained in the introduction, this provides a QFT definition for
$\varep$ (in general, different for the bulk and boundary cases),
hence a universal definition for the finite terms in $S_A$.

The general definition of the branch-point twist fields as
explained in subsection \ref{sspartfunct} did not say anything
about the normalisation of the fields. This normalisation does not
depend on the positions of the fields, but may, and in general
does, depend on the number of copies $n$ -- that is, a change of
normalisation may change the $n$-dependent number $Z_n$ in
(\ref{rhon}) and (\ref{be3}). Let us choose the {\em CFT
normalisation}: this is the normalisation by which the
short-distance behaviour of the two-point function is exactly
given by the CFT two-point function, with coefficient 1:
\beq\label{CFTn}
    \bra0| \tw(x_1) \t\tw(x_2)|0\ket \sim |x_1-x_2|^{-2d_n}\quad \mbox{as}\quad m|x_1-x_2|\to0.
\eeq This two-point function just depends on the euclidean
distance $|x_1-x_2|$, thanks to euclidean translation and rotation
invariance (or Poincar\'e invariance in Minkowsky space), and to
the fact that the twist fields are spinless. With this
normalisation, and with a fixed relation between $\varep$ and the
correlation length $\xi$, we have a fixed function $Z_n$ in
(\ref{result}). In particular, $Z_1=1$, because with the CFT
normalisation, the twist fields just become the identity operator
${\bf 1}$ at $n=1$. We may change the definition of $\varep$ by
rescaling it in an $n$-independent way, so that we change its
relation to the correlation length, but keep it independent from
$n$. This rescaling is absorbed into $Z_n$, and we may do so in
order that $d Z_n/dn = 0$ at $n=1$, as said in the introduction.
With this rescaling of $\varep$, it is easy to see that taking the
derivative in (\ref{rhon}) with the short-distance behaviour
(\ref{CFTn}), we immediately obtain the UV asymptotic given by
(\ref{shlabu}) for the entanglement entropy.

Concerning the boundary case, we need to choose the normalisation
of  the boundary state $|B\ket$ in order that with the CFT
normalisation (\ref{CFTn}), the expression (\ref{be3}) gives rise
to the correct IR asymptotic in (\ref{shla}). In that IR
asymptotic, there is the constant $U^{\text{model}}$ that appears,
a constant that is defined in the bulk situation. At large $r$,
the two-point function in the bulk factorises into a product of
one-point functions (this is asymptotic factorisation of
correlation functions of local fields): \beq
    \bra0| \tw(x_1) \t\tw(x_2)|0\ket \sim (\bra0|\tw|0\ket)^2\quad \mbox{as}\quad m|x_1-x_2|\to\infty
\eeq where we used the fact that $\t\tw$ is obtained from $\tw$ by
a permutation symmetry transformation that keeps the vacuum state
invariant. Using this into (\ref{rhon}), we find the correct IR
behaviour of (\ref{shlabu}), with \beq\label{Ugen}
    U^{\text{model}} = -2\lim_{n\to1} \frc{d}{dn} \lt(m^{-d_n} \bra0|\tw|0\ket\rt).
\eeq Hence, this constant is related to the one-point function of
twist fields in the CFT normalisation.  Returning to the boundary
case, with the choice of boundary state normalisation
\beq\label{bsn}
    \bra 0|B\ket = 1,
\eeq
we see that a decomposition into energy and momentum eigenstates, keeping the ground state only, gives
\beq
    \bra 0|\tw(x)|B\ket \sim \bra 0|\tw|0\ket \quad \mbox{as}\quad mr|\to\infty
\eeq where $r$ is the distance between $x$ and the boundary.
Hence, we indeed find, from (\ref{be3}), the correct IR behaviour
given by (\ref{shla}) for the entanglement entropy in the boundary
case.

With this normalisation, the model-dependent and boundary-condition-dependent constant $V(\kappa)$ is then universal. Since it occurs at small distances, it is derived from the UV behaviour of the one-point function in (\ref{be3}). This behaviour is itself completely controlled by the conformal UV fixed point, hence at short distances that one-point function is a CFT one-point function. This means that the constant $V(\kappa)$ only depends on the conformal boundary condition to which the system flows as $mr\to0$. This was seen explicitly in the Ising model in \cite{nexttonext}, which will be reviewed in section \ref{sectent}.

\sect{The form factor program for branch-point twist fields}

We will now describe how to evaluate correlation functions of
branch-point  twist fields, involved in the universal part of the
entanglement entropy in the bulk and boundary cases through the
expressions (\ref{rhon}) and (\ref{be3}). The method we will use
is that of form factor expansion; this is a representation of the
two-point function (in the bulk case) or the one-point function
(in the boundary case) that uses the decomposition of the identity
operator into projection operators on energy and momentum
eigenstates. It is an effective large-distance (large-$r$)
expansion, and in principle, its re-summation gives a
representation valid at all non-zero distances.

\subsection{Form factor expansion in 1+1-dimensional massive
QFT}\label{ssectffexp}

We turn to the description of QFT on Minkowski space-time in terms
of its Hilbert space of asymptotic relativistic particles. This
will allow us to introduce the  method we use in order to evaluate
the correlation functions of branch-point twist fields: the form
factor expansion.

In the context of $1+1$-dimensional QFT, form factors are defined
as tensor valued functions representing matrix elements of some
local operator $\mathcal{O}(x)$ located at the origin $x=0$
between a multi-particle {\em{in}}-state and the vacuum:
\begin{equation}
F_{k}^{\mathcal{O}|\mu _{1}\ldots \mu _{k}}(\theta _{1},\ldots
,\theta _{k}):=\left\langle
0|\mathcal{O}(0)|\theta_1,\ldots,\theta_k\right\rangle_{\mu_1,\ldots,\mu_k}^{\text{in}}
~.\label{ff}
\end{equation}
Here $|\theta_1,\ldots,\theta_k\rangle_{\mu_1,\ldots,\mu_k}^{\text{in}}$ represent
the physical ``in'' asymptotic states of massive QFT. They carry
indices $\mu_i$, which are quantum numbers characterizing the
various particle species, and depend on the real parameters
$\theta_i$, which are the associated rapidities. The form factors are
defined for all rapidities by analytically continuing from some
ordering of the rapidities; a fixed ordering provides a complete
basis of states, for instance $\theta_1>\theta_2>\cdots>\theta_k$. Form factors also depend on the number of copies $n$,
but for simplicity we will not write this dependence explicitly. Moreover, because of
relativistic invariance and spinlessness of the branch-point twist fields, they in fact only depend on the rapidity differences; in the two-particle case, they then become functions of only one variable, $\theta=\theta_1-\theta_2$.

The twist property associated to the permutation symmetry in the
general definition of branch-point twist fields was explained in
subsection \ref{sspartfunct}. It is at the basis of the main
properties of the form factors of branch-point twist fields, and
ultimately of our large-distance result about the entanglement
entropy. It implies the following fundamental exchange relations
for branch-point twist fields as operators on the space of
asymptotic states. If $\Psi_{1}, \ldots, \Psi_{n}$ are the fields
associated to the fundamental particles of each copy of the
original model, then the equal time ($x^0=y^0$) exchange relations
between $\mathcal{T}$ and $\Psi_{1}, \ldots, \Psi_{n}$ can be
written in the following form\footnote{Here we employ the standard
notation in Minkowski space-time: $x^{\nu}$ with $\nu=0,1$, with
$x^{0}$ being the time coordinate and $x^{1}$ being the position
coordinate.}
\begin{eqnarray}
    \Psi_{i}(y)\mathcal{T}(x) &=& \mathcal{T}(x) \Psi_{i+1}(y) \qquad x^{1}> y^{1}, \n
    \Psi_{i}(y)\mathcal{T}(x) &=& \mathcal{T}(x) \Psi_{i}(y) \qquad x^{1}< y^{1}, \label{cr}
\end{eqnarray}
for $i=1,\ldots, n$ and where we identify the indices $n+i \equiv
i$. The twist field $\tilde{\mathcal{T}}$ has similar properties
as $\mathcal{T}$, with the difference that its exchange relations
with the fundamental fields of the theory are given by
\begin{eqnarray}
    \Psi_{i}(y)\tilde{\mathcal{T}}(x) &=& \tilde{ \mathcal{T}}(x) \Psi_{i-1}(y) \qquad x^{1}> y^{1}, \n
    \Psi_{i}(y)\tilde{\mathcal{T}}(x) &=& \tilde{ \mathcal{T}}(x) \Psi_{i}(y) \qquad x^{1}< y^{1}, \label{cr2}
\end{eqnarray}
instead of (\ref{cr}). The connection with the previous section is obtained by
recalling that in going from the Hilbert space description to the
path integral description, the order of operators is translated
into time-ordering (or $\ry$-ordering in euclidean space), and
that left-most operators are later in time.

For our present purpose, what is important is that these exchange
relations, along with the uniqueness of the branch-point twist
fields explained in subsection \ref{sspartfunct}, give us the
following relation between branch-point twist fields, as operators
on the space of asymptotic states: \beq\label{hc}
    \t\tw = \tw^\dag.
\eeq

In order to obtain large-distance expansions, we simply introduce a complete sum over quantum states
\begin{equation}
   {\bf 1}=\sum_{k=1}^{\infty
}\sum_{\mu _{1}\ldots \mu _{k}}\int\limits_{\theta_1>\theta_2>\cdots>\theta_k }\frac{d\theta _{1}\ldots d\theta _{k}}{%
(2\pi
)^{k}}|\theta_1,\ldots,\theta_k\rangle_{\mu_1,\ldots,\mu_k}\,
   {\,}_{\mu_k,\ldots,\mu_1}\!\langle
   \theta_k,\ldots,\theta_1|\label{csum}
\end{equation}
The two-point function in (\ref{rhon}) is then expressed as
\beq\label{bulkexp}
    \bra0| \tw(x_1) \t\tw(x_2) |0\ket
    = \sum_{k=1}^{\infty
}\sum_{\mu _{1}\ldots \mu _{k}}\int\limits_{\theta_1>\theta_2>\cdots>\theta_k }\frac{d\theta _{1}\ldots d\theta _{k}}{%
(2\pi )^{k}} e^{-r\sum\limits_{j=1}^k m_{\mu_j} \cosh\theta_j}
|\bra
0|\tw(0)|\theta_1,\ldots,\theta_k\rangle_{\mu_1,\ldots,\mu_k}|^2
\eeq where $r=|x_1-x_2|$ is a space-like distance between the
points $x_1$ and $x_2$ in Minkowsky space-time. Here, we used the
fact that $\t\tw$ is the hermitian conjugate of $\tw$ (\ref{hc}),
and we used translation covariance in space and time in order to
extract the exponential factor containing all the position
dependence. We also used relativistic invariance and spinlessness
to say that the form factors only depend on the rapidity
differences. In this way, we could do a simple shift of rapidities
in the imaginary direction, so that only $\cosh\theta_j$'s are
left in the exponential, with coefficients the negative of the
space-like relativistic distance times the mass of particle
$\mu_j$. The integrals, then, become explicitly exponentially
convergent (the large-rapidity asymptotics of the form factors is
at most exponential in $\theta_j$'s). In the same spirit, we could
have simply thought of $x_1$ and $x_2$ as coordinates in $\R^2$
(with imaginary time), used the euclidean rotation invariance in
order to bring $x_1-x_2$ in the pure imaginary time direction
(with a distance $r$), then used imaginary-time translation
covariance to extract the real-exponential factor depending on
$r$. The factor $1/(2\pi)^k$ in the expansion just tells us about
the normalisation we have chosen for the asymptotic states.

Similarly, the one-point function in (\ref{be3}) is expressed as
\begin{eqnarray}\label{boundexp}
    \langle 0| {{\cal T}}(x)|B\rangle
   & =& \sum_{k=1}^{\infty
}\sum_{\mu _{1}\ldots \mu _{k}}\int\limits_{\theta_1>\theta_2>\cdots>\theta_k }\left[\frac{d\theta _{1}\ldots d\theta _{k}}{%
(2\pi )^{k}} \,\,  e^{-r\sum\limits_{j=1}^k m_{\mu_j}
\cosh\theta_j}\right.
\nonumber\\
&& \left.\times\bra
0|\tw(0)|\theta_1,\ldots,\theta_k\rangle_{\mu_1,\ldots,\mu_k}
    {\,}_{\mu_k,\ldots,\mu_1}\!\langle
   \theta_k,\ldots,\theta_1|B\ket\right].
\end{eqnarray} Here, the easiest way to obtain this expansion is by
considering that the distance $r$ between $x$ and the boundary is
in the imaginary time direction, and using imaginary-time
translation covariance in order to bring out the real-exponential
factor.

In the expansions (\ref{bulkexp}) and (\ref{boundexp}) the terms
with a larger number of particles are smaller because the lowest
possible value of the argument of the exponential is $-r k m_{1}$,
$k$ times the lowest mass $m_1$. Moreover, as $r$ increases, the
larger-particle terms decrease faster; hence, these indeed are
large-distance expansions.

\subsection{Form factor equations in integrable QFT}

We have found that (\ref{CFTn}) and (\ref{bsn}) provides the
correct  field and boundary state normalisations in order to
reproduce (\ref{shlabu}) and (\ref{shla}). We have also explained
how the form factors can be used in order to calculate the
correlation functions involved in the universal part of the
entanglement entropy. Let us now provide more explanation as to
the properties of form factors in the context of integrable models
of QFT.

The main characteristics of massive integrable models of QFT are
that the number of particles and their  momenta set are conserved
under scattering, and that the scattering matrix factorises into
products of two-particle scattering matrices. The two-particle scattering matrix  (or $S$-matrix) is the
solution of a set of consistency equations and analytic
properties. These consistency equations and analytic properties
are often strong enough, when combined with general properties of the model (symmetries, etc.), to completely fix the $S$-matrix in
integrable models. In a similar way, the form factors can be completely fixed by solving a set
of equations and analytic properties, which now depend on the two-particle
$S$-matrix.

In this section we want to show how the standard form factor
equations for $1+1$-dimensional IQFTs must be modified for the
branch-point twist fields. Here we will consider an integrable
model consisting of $n$ copies of a known integrable theory
possessing a single particle spectrum and no bound states (such as
the Ising and sinh-Gordon models). The interested reader may refer
to \cite{next} for the general case. We have therefore $n$
particles, which we will denote by indices $1, \ldots, n$. The
$S$-matrix between particles $i$ and $j$ with rapidities
$\theta_i$ and $\theta_j$ will be denoted by
$S_{ij}(\theta_i-\theta_j)$ (that it depends on the rapidity
difference is a consequence of relativistic invariance). Particles
of different copies do not interact with each other, so that the
$S$-matrix of the model will be of the form
\begin{eqnarray}\label{s}
   S_{ij}(\theta)&=& S(\theta)^{\delta_{ij}} \qquad \forall \quad i,j=1, \ldots,
   n,
\end{eqnarray}
where  $S(\theta)$ is the $S$-matrix of the single-copy integrable
QFT.

\begin{figure}
\bc
\includegraphics[width=6cm,height=5cm]{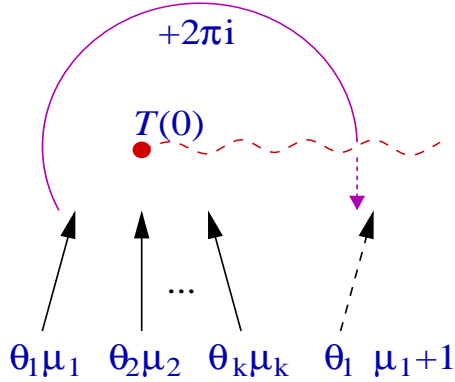}
\ec \caption{A pictorial representation of the effect of adding
$2\pi i$ to rapidity $\theta_1$ in form factors of the twist field
$\tw$.} \label{fig-periodicity}
\end{figure}

\begin{figure}
\bc
\includegraphics[width=14cm,height=5cm]{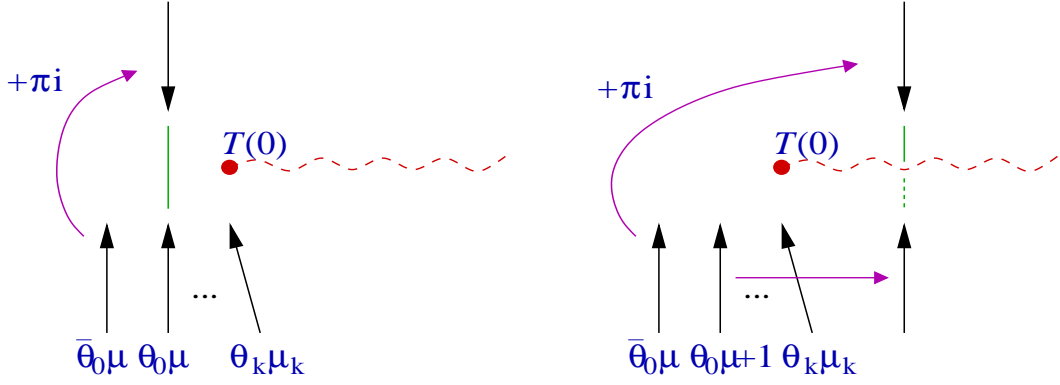}
\ec \caption{The kinematic poles come from the structure of the
wave function far from the local fields, at positive and negative
infinity. Adding $i\pi$ to rapidity $\theta_1$ puts the particle
in the ``out" region. With a particle in that region, there are
delta-functions representing particles in the ``in" region going
through without interacting with the local fields. Those occur
from the $e^{ipx}$ form of the wave function at positive and
negative infinity. But if the coefficients at both limits are
different, $S_- e^{ipx}$ and $S_+e^{ipx}$ with $S_-\neq S_+$, then
there are also poles in addition to these delta-functions. Only
these poles are seen in the analytic continuation
$\theta_1\mapsto\theta_1+i\pi$. Different coefficients come from
the semi-locality of the twist field and the non-free scattering
matrix, as represented here.} \label{fig-kinematic}
\end{figure}

It is well known that exchange relations like (\ref{cr}) and (\ref{cr2}) play an important role in the derivation
of the consistency equations for the form factors. Generalising the standard arguments to the exchange
relation (\ref{cr}), the form factor axioms are
\begin{eqnarray}
  F_{k}^{\mathcal{T}|\ldots \mu_i  \mu_{i+1} \ldots }(\ldots,\theta_i, \theta_{i+1}, \ldots ) &=&
  S_{\mu_i \mu_{i+1}}(\theta_{i\,i+1})
  F_{k}^{\mathcal{T}|\ldots \mu_{i+1}  \mu_{i} \ldots}(\ldots,\theta_{i+1}, \theta_i,  \ldots ),
  \label{1}\\
 F_{k}^{\mathcal{T}|\mu_1 \mu_2 \ldots \mu_k}(\theta_1+2 \pi i, \ldots,
\theta_k) &=&
  F_{k}^{\mathcal{T}| \mu_2 \ldots \mu_n \hat{\mu}_1}(\theta_2, \ldots, \theta_{k},
  \theta_1),\label{2}\\
 \begin{array}{l}
\\
  \text{Res}  \\
 {\footnotesize \bar{\theta}_{0}={\theta}_{0}}
\end{array}\!\!\!\!
 F_{k+2}^{\mathcal{T}|\b{\mu} \mu  \mu_1 \ldots \mu_k}(\bar{\theta}_0+i\pi,{\theta}_{0}, \theta_1 \ldots, \theta_k)
  &=&
  i \,F_{k}^{\mathcal{T}| \mu_1 \ldots \mu_k}(\theta_1, \ldots,\theta_k), \label{3}
  \\
\begin{array}{l}
\\
  \text{Res}  \\
 {\footnotesize \bar{\theta}_{0}={\theta}_{0}}
\end{array}\!\!\!\!
 F_{k+2}^{\mathcal{T}|\b\mu \hat{\mu } \mu_1 \ldots \mu_k}(\bar{\theta}_0+i\pi,{\theta}_{0}, \theta_1 \ldots, \theta_k)
  &=&-i\prod_{i=1}^{k} S_{\hat{\mu}\mu_i}(\theta_{0i})
  F_{k}^{\mathcal{T}| \mu_1 \ldots \mu_k}(\theta_1, \ldots,\theta_k).\label{kre}
\end{eqnarray}
Here $\theta_{ij}=\theta_i-\theta_j$. Besides the simple poles whose residues are given in (\ref{3}) and (\ref{kre}), the form factors are, as function of $\theta_{ij}$, analytic in the strip ${\rm Im}(\theta_{ij}) \in [0,2\pi)$ (there would be additional poles, with prescribed residues, if there were bound states; but we only consider integrable models without bound states here). The first axiom is in
fact the same as for ordinary local fields. In the second equation, the
crossing or locality relation, we introduced the symbols
$\hat{\mu}_i=\mu_i+1$. As compared to the usual form factor
equations, it is altered by the nature of the exchange relation
and it now relates form factors associated to different particle
sets. Finally, the last two equations generalise the standard
kinematic residue equation to branch-point twist fields. Once
more, the exchange relations (\ref{cr}) are responsible for the
splitting into two equations. Here, for later convenience, we
wrote the equations in their general form valid also for
many-particle models, where $\b\mu$ represents the anti-particle
associated to $\mu$. In the present case, the integrable model we
started with has just one particle (so that $\mu$ labels the
copies) and therefore each particle is its own anti-particle.
Pictorial
explanations of the second and of the last two equations are
given, respectively, in Figs. \ref{fig-periodicity} and
\ref{fig-kinematic}.

\subsection{Form factor equations in non-integrable 1+1-dimensional
QFT} \label{ssectni}

 Out of integrability, we do not have the luxury of the large
simplification of the scattering amplitudes. However, two-particle
form factors still satisfy certain properties. For simplicity, we
will keep the notation as if there were only one particle type in
the original model, so that $\mu$ represents the copy number, but
attaching an extra index to $\mu$ does not affect any of the
considerations or equations below.

In the case of an ordinary, spinless local field $\Or$,  the
two-particle form factor $F_2^{\Or|\mu_1\mu_2}(\theta_1,\theta_2)$
is a function of $\theta=\theta_1-\theta_2$ with certain
properties when considered as an analytic function of $\theta$,
analytically continued from $\theta>0$ where it is the form factor
with ``in'' asymptotic states as defined above. The properties of
this function are:
\begin{itemize}
\item the function analytically continued to negative  rapidities
$\theta<0$ along the real line is a form factor with an ``out''
asymptotic state: \begin{equation}
    F_2^{\Or|\mu_2\mu_1}(\theta_2-\theta_1) \stackrel{\theta_1>\theta_2}=
    \left\langle
    0|\mathcal{O}(0)|\theta_1,\theta_2\right\rangle_{\mu_1,\mu_2}^{\text{out}}.\label{ni1gen}
\end{equation}
 \item the function analytically continued by a $2\pi i$ shift
from the real line gives: \beq
    F_2^{\Or|\mu_1\mu_2}(\theta_1-\theta_2+2\pi i) = F_2^{\Or|\mu_2\mu_1}(\theta_2-\theta_1)
\eeq \item the function is analytic in the physical strip ${\rm
Im}(\theta)\in[0,2\pi)$ except possibly for poles on ${\rm
Re}(\theta)=0$, ${\rm Im}(\theta)\in(0,\pi)\cup (\pi,2\pi)$
corresponding to bound states.
\end{itemize}
Note that the physical strip in the rapidity plane,  in the last
point, is a double covering of the physical sheet in the
Mandelstam's $s$-plane, obtained through $s = m_1^2+m_2^2+2m_1m_2
\cosh(\theta_1-\theta_2)$. This double covering has a symmetry
under flip with respect to the point $i\pi$, as expressed by the
second point above. For instance, for ${\rm Im}(\theta)=2\pi$ and
${\rm Re}(\theta)>0$, the function is describing a form factor
with an ``out'' asymptotic state.

The analyticity conditions above are somewhat strong  conditions,
but they are very far from allowing us to fix the two-particle
form factors, because in general, the analytic structure outside
of the physical strip is very complicated. However, in the case of
the branch-point twist fields, the generalisation of these
conditions are strong enough to give a result for the entanglement
entropy. The generalisation is obtained through arguments similar
to those presented in the case of integrable models, whereby wave
packets are exchanged or rotated around the point where the twist
field lies, keeping only in mind that the exchange of wave packets
does not lead to a simple two-particle $S$-matrix in the
non-integrable case. The first point above is unchanged by the
twist property, as was equation (\ref{1}) in the case of
integrable models, because it involves wave packet exchanges only
on one copy of the replica model. Since in the replica model,
particles on different copies do not interact even in general,
non-integrable QFT, we still have here (\ref{1}) in the
two-particle case when the copy numbers are different:
\beq\label{ni1}
    F_2^{\tw|\mu_2\mu_1}(-\theta) = F_2^{\tw|\mu_1\mu_2}(\theta) \quad (\mu_1\neq\mu_2).
\eeq The second point above is affected by the twist property, in
exactly the same way as equation (\ref{2}) was in the integrable
case. Hence, equation (\ref{2}) still holds in the non-integrable,
two-particle case: \beq\label{ni2}
    F_2^{\tw|\mu_1\mu_2}(\theta+2\pi i) = F_2^{\tw|\mu_2\h\mu_1}(-\theta)
\eeq where we recall that $\h\mu = \mu+1$. Finally, the third
point above holds,  up to one additional pole, the kinematic pole,
as in the integrable case, equations (\ref{3}) and (\ref{kre}).
Indeed, the argument for this kinematic pole to be present, given
in Fig. \ref{fig-kinematic}, holds true when there is only two
particles in the non-integrable case, because there is no exchange
of wave packets necessary. Hence, we have
\begin{eqnarray}
 \begin{array}{l}
\\
  \text{Res}  \\
 {\footnotesize \theta=0}
\end{array}\!\!\!\!
 F_{2}^{\mathcal{T}|\b{\mu} \mu }(\theta+i\pi)
  &=&
  i \,\bra0|\tw|0\ket, \label{ni3}
  \\
\begin{array}{l}
\\
  \text{Res}  \\
 {\footnotesize \theta=0}
\end{array}\!\!\!\!
 F_{2}^{\mathcal{T}|\b\mu \hat{\mu } }(\theta+i\pi)
  &=&-i\,\bra0|\tw|0\ket. \label{nikre}
\end{eqnarray}

Equations (\ref{ni1}), (\ref{ni2}), (\ref{ni3}) and (\ref{nikre})
are what is left of (\ref{1}), (\ref{2}), (\ref{3}) and
(\ref{kre}) when integrability is taken away. As we will see
below, in the integrable case, and only in this case, we can write
exact form factors, but in general, the equations left in the
non-integrable case are enough to obtain the exact result
displayed in the introduction.

\subsection{Two-particle form factors}

\begin{figure}
\bc
\includegraphics[width=6cm,height=4cm]{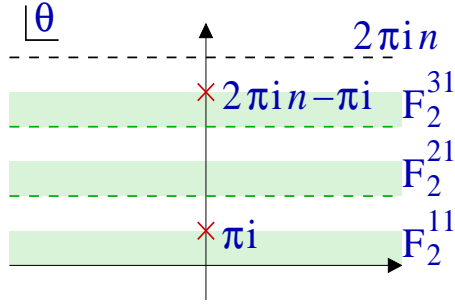}
\ec \caption{The structure of the function $F_2^{\tw|11}(\theta)$
in the extended sheet ${\rm Im}(\theta)\in[0,2\pi n]$, in the case
$n=3$. Crosses indicate the positions of the kinematic
singularities. Shaded regions represent the physical sheets of the
form factors $F_2^{\tw|11}(\theta)$, $F_2^{\tw|12}(\theta)$ and
$F_2^{\tw|13}(\theta)$.} \label{fig-structure}
\end{figure}

Let us now analyse the equations for two-particle form
factors  in more detail. First, it is clear that by a cyclic
exchange of the sheets, the branch-point twist fields are
unchanged, so that form factors are invariant under such a cyclic
exchange. Hence, it is sufficient to look at
$F^{\tw|j1}_2(\theta)$, for instance. Second, we may repeatedly
use (\ref{ni2}) and (\ref{ni1}) in order to obtain
$F^{\tw|j1}_2(\theta)$ for $j=2,3,\ldots,n$. For instance:
\[
    F^{\tw|11}_2(\theta+2\pi i) = F^{\tw|12}_2(-\theta) = F^{\tw|21}_2(\theta),\quad
    F^{\tw|21}_2(\theta+2\pi i) = F^{\tw|13}_2(-\theta) = F^{\tw|31}_2(\theta),
\]
etc. Hence, we find
\begin{eqnarray}
  F_{2}^{\mathcal{T}|i \, i+k}(\theta) &=&F_{\text{min}}^{\mathcal{T}|j
  \,
  j+k}(\theta) \qquad
\forall \quad i,j,k\label{use}\\
F_{2}^{\mathcal{T}|
j1}(\theta)&=&F_{\text{min}}^{\mathcal{T}|11}(\theta+2\pi
i(j-1))\qquad j=2,3,\ldots,n.\label{use2}
\end{eqnarray}
Finally, the residue equations (\ref{ni3}) and (\ref{nikre}) as well as the analyticity conditions on the physical strip tell us about the analytic structure of $F_2^{\tw|11}(\theta)$ in the larger region ${\rm Im}(\theta) \in[0,2\pi i n)$, which we call the {\em extended physical strip} \cite{entropy}. The function $F_2^{\tw|11}(\theta)$ is analytic everywhere in that strip, except for simple poles on ${\rm Re}(\theta)=0$: at $\theta=i\pi$, with residue $i\bra0|\tw|0\ket$, and $\theta=2i\pi n-i\pi$, with residue $-i\bra 0|\tw|0\ket$, and, possibly, on $0<{\rm Im}(\theta)<\pi$ and $2\pi n-\pi<{\rm Im}(\theta)<2\pi n$ if there are bound states. This structure is depicted in Fig. \ref{fig-structure} for the case $n=3$ and without bound states.

This structure is all we can say about the two-particle form factors in the non-integrable case. In the integrable case, however, we can say much more; again, we concentrate here on diagonal scattering with a one-particle spectrum and without bound states for simplicity. As usual in the integrable context, we define the minimal form
factors $F_{\text{min}}^{\mathcal{T}|j k}(\theta, n)$ to be
solutions of the equations (\ref{1}) and (\ref{2}) for $k=2$
without poles in the physical strip ${\rm Im}(\theta) \in
[0,2\pi]$. That is,
\begin{eqnarray}
F_{\text{min}}^{\mathcal{T}|kj}(\theta)=F_{\text{min}}^{\mathcal{T}|jk}(-\theta)S_{kj}(\theta)=
F_{\text{min}}^{\mathcal{T}|j\, k+1}(2\pi i-\theta) \qquad \forall
\quad j,k \label{mini}
\end{eqnarray}
where the $S$-matrix is given by (\ref{s}). Minimal form factors satisfy the same constraints (\ref{use}) and (\ref{use2}), a consequence of which is that the minimal form factor $F_{\text{min}}^{\mathcal{T}|11}(\theta)$ must have no
poles in the strip $\text{Im}(\theta) \in [0, 2\pi n]$. From the equations above it is easy to deduce
\begin{equation}
  F_{\text{min}}^{\mathcal{T}|11}(\theta)=F_{\text{min}}^{\mathcal{T}|11}(-\theta)S(\theta)=
  F_{\text{min}}^{\mathcal{T}|11}(-\theta+ 2\pi n i). \label{rel}
\end{equation}
These are the main equations from which we can determine all minimal form factors in integrable models.

In order to develop a systematic procedure to solve these equations, it is useful to recall that, for a standard local
operator the minimal form factor equations take the form
\begin{equation}
  f_{11}(\theta)=f_{11}(-\theta) {S}(n\theta)= f_{11}(-\theta+ 2\pi
  i),
   \label{rel2}
\end{equation}
provided that the $S$-matrix of the theory is given by
$S(n\theta)$. Thus given a solution to the previous equation, the function
$F_{\text{min}}^{\mathcal{T}|11}(\theta, n)=f_{11}(\theta/n)$ is
automatically a solution of (\ref{rel}).

In the context of integrable models, a systematic way
of solving such type of equations has been developed whereby,
given an integral representation for $S(\theta)$, an integral
representation of $f_{11}(\theta)$ can be readily obtained
\cite{KW}. For diagonal theories, the integral representation of
the $S$-matrix takes the form
\begin{equation}
    S(\theta)=\exp \left[\int_{0}^{\infty} \frac{dt }{t} g(t) \sinh\left(\frac{t \theta}{i
    \pi}\right)\right], \label{irep}
\end{equation}
where $g(\theta)$ is a function which depends of the theory under
consideration. A trivial consequence of the previous equation is
\begin{equation}
    S(n \theta)=\exp\left[ \int_{0}^{\infty} \frac{dt }{t} g(t/n) \sinh\left(\frac{t \theta}{i
    \pi}\right)\right],
\end{equation}
and from here, it is easy to show that
\begin{equation}
    f_{11}(\theta)=\mathcal{N} \exp\left[\int_{0}^{\infty} \frac{dt}{t \sinh(n t)} g(t) \sin^{2}\left(\frac{i t n}{2}\left(1+\frac{i\theta}{\pi}\right)
    \right)\right]
\end{equation}
where $\mathcal{N}$ is a normalization constant. Therefore, the
desired solution is
\begin{equation}
    F_{\text{min}}^{\mathcal{T}|11}(\theta)=
    f_{11}(\theta/n)=\mathcal{N} \exp\left[\int_{0}^{\infty} \frac{dt}{t \sinh(n t)} g(t) \sin^{2}\left(\frac{i t}{2}\left(n+\frac{i\theta}{\pi}\right)
    \right)\right].
\end{equation}
In order to evaluate the full two-particle form factors, we must solve (\ref{mini}) with the poles
in the extended physical strip mentioned before. The ``minimal'' solution, with in particular the least diverging behaviour at large $\theta$, is
\begin{equation}
F_{2}^{\mathcal{T}|jk}(\theta)=\frac{ \langle \mathcal{T}\rangle
\sin\left(\frac{\pi}{n}\right)}{2 n \sinh\left(\frac{ i\pi
(2(j-k)-1)+ \theta}{2n}\right)\sinh\left(\frac{i\pi
(2(k-j)-1)-\theta}{2n}\right)}
\frac{F_{\text{min}}^{\mathcal{T}|jk}(\theta)}{F_{\text{min}}^{\mathcal{T}|jk}(i
\pi)},\label{full}
\end{equation}
Note that the constant factor $\sinh(\pi/n)$ guarantees that
all form factors vanish for $n=1$ as expected, since in that case
the field $\mathcal{T}$ can be identified with the identity.

For the field $\t\tw$, the exchange relations imply that form
factors of the field $\tilde{\mathcal{T}}$ are equal to those of
the field $\mathcal{T}$ up to the transformation $i \rightarrow
n-i$ for each particle $i$. This means that
\begin{equation}
    F_2^{\mathcal{T}|ij}(\theta)=F_2^{\tilde{\mathcal{T}}|(n-i)
    (n-j)}(\theta).
\end{equation}
This property can be combined with (\ref{use})-(\ref{use2}) to
show that
\begin{eqnarray}
   F_2^{\tilde{\mathcal{T}}|11}(\theta)&=&F_2^{\mathcal{T}|11}(\theta).\\
  F_2^{\tilde{\mathcal{T}}|1j}(\theta)&=&F_2^{\mathcal{T}|11}(\theta+2\pi i(j-1)).
\end{eqnarray}

\subsubsection{Examples: the Ising and sinh-Gordon models}

The Ising and sinh-Gordon models are examples of the
kind of integrable models for which we have specialised the
results of this section: they have a single particle spectrum and
no bound states. Their two-particle $S$-matrices are,
\begin{equation}
S_{\text{I}}(\theta)=-1,\quad\quad\text{and}\quad\quad
S_{\text{s-G}}(\theta)=\frac{\tanh\frac{1}{2}(\theta-i \frac{\pi
B}{2}) }{\tanh \frac{1}{2}(\theta+i \frac{\pi
    B}{2})},\label{smatrices}
\end{equation}
respectively. In the sinh-Gordon model, the parameter $B \in
[0,2]$ is the effective coupling constant which is related to the
coupling constant $\beta$ in the sinh-Gordon Lagrangian
\cite{toda2,toda1} as
\begin{equation}\label{BB}
    B(\beta)=\frac{2\beta^2}{8\pi + \beta^2},
\end{equation}
under CFT normalization \cite{za}. The $S$-matrix
\cite{SSG2,SSG3,SSG} is obviously invariant under the
transformation $B\rightarrow 2-B$, a symmetry which is also
referred to as week-strong coupling duality, as it corresponds to
$B(\beta)\rightarrow B(8 \pi \beta^{-1})$ in (\ref{BB}). The point
$B=1$ is known as the self-dual point.

Form factors of local fields of the Ising model were first
computed in \cite{KW,YZam} and later on in \cite{Cardybast} for
so-called descendant fields. A solution of (\ref{mini}) for
$j=k=1$ is given by
\begin{equation}
F_{\text{min}}^{\mathcal{T}|11}(\theta)=-i\sinh\left(\frac{\theta}{2n}\right),
\quad \text{for the Ising field theory}.\label{is}
\end{equation}
In accordance with our previous arguments, this is the standard minimal form factor already employed
in \cite{KW,YZam}, with $\theta \rightarrow\theta/n$.

Form factors of the sinh-Gordon model were first computed in
\cite{FMS}. The program was thereafter extended to other operators
in \cite{KK} and more recently in \cite{Delfino:2006te}. The
$S$-matrix above admits an integral representation which is given
by (\ref{irep}), with
\begin{equation}\label{ft}
    g(t)=\frac{8\sinh\left( \frac{tB}{4}\right) \sinh\left(\frac{t}{2}\left(1-\frac{B}{2} \right)
    \right)\sinh\left( \frac{t}{2}\right)}{\sinh t}.
\end{equation}
Therefore, the minimal form factor is given by
\begin{equation}
 F_{\text{min}}^{\mathcal{T}|11}(\theta)=\exp\left[-2
 \int_{0}^{\infty} \frac{dt \sinh \frac{t B}{4} \sinh  \frac{t(2-B)}{4} \cosh t\left(n+\frac{i\theta}{\pi}
    \right)}{t \sinh(n t)
 \cosh \frac{t}{2} }\right]\quad \text{for the sinh-Gordon model},\label{int}
\end{equation}
where we have chosen the normalization
$\mathcal{N}=F_{\text{min}}^{\mathcal{T}|11}(i \pi n)$. This
integral representation can be turned into an infinite product of
ratios of Gamma-functions as described in \cite{entropy}. As a
consistency check, it is quite easy to show that for $n=1$ the
minimal form factor above is the standard minimal form factor
associated to local fields in the sinh-Gordon model computed in
\cite{FMS}. In appendix A of \cite{entropy} we showed how the same
expression can be derived from the angular quantization scheme
\cite{freefield1,freefield2}. The branch-point twist field minimal form factors
were also computed for the sine-Gordon model in \cite{other}.

\subsection{Higher particle form factors}\label{higher}

Form factors with a number of particles higher than 2 can
only  be dealt with in the integrable case. From the consistency
equations for form factors of the twist field $\mathcal{T}$ and
generalizing the methods previously used for the computation of
form factors of local fields in diagonal theories (see e.g.
\cite{FMS}), it is natural to make the following ansatz for the
higher particle form factors:
\begin{equation}\label{hp}
    F_k^{\mathcal{T}|11\cdots 1} (\theta_1, \ldots, \theta_k)=H_{k}^{\mathcal{T}|1\ldots 1}Q_{k}^{\mathcal{T}|1\ldots 1}(x_1, \ldots,
    x_k)\prod_{i
    <j}{K(\theta_{ij})},
\end{equation}
where
\begin{equation}\label{111}
K(\theta_{ij})=
{F}_{2}^{\mathcal{T}|11}(\theta_{ij})/\langle\mathcal{T}\rangle,
\end{equation}
and we are considering form factors of $k$ particles of the same
type (say type 1), since any other form factors can be obtained
from these by repeated use of (\ref{1}) and (\ref{2}). More
precisely,
\begin{equation}
    F_k^{\mathcal{T}|\mu_1\cdots \mu_k} (\theta_1, \ldots,
    \theta_k)= F_k^{\mathcal{T}|11\cdots 1} (\theta_1+ 2\pi i (\mu_1-1), \theta_2+2\pi
i (\mu_2-1),\ldots,
    \theta_k+2 \pi i (\mu_k-1)), \label{tralala}
\end{equation}
with the ordering $\mu_1\geq \mu_2 \geq \ldots \geq \mu_k$. It is
convenient to introduce the variables $x_i=\exp(\theta_i/n)$ with
$i=1, \ldots, k$. Then, exploiting the properties of the
two-particle form factor discussed above, we find that the first
two form factor consistency equations (\ref{1}) and (\ref{2}) are
automatically satisfied, provided that the functions
$Q_{k}^{\mathcal{T}|1\ldots 1}(x_1, \ldots, x_k)$ are symmetric in
all variables $x_i$. In other words, $Q_{k}^{\mathcal{T}|1\ldots
1}(x_1, \ldots, x_k)$ can be expressed in terms of elementary
symmetric polynomials of the variables $x_i$.
$H_{k}^{\mathcal{T}|1\ldots 1}$ are constants. In particular,
$H_0^{\mathcal{T}}=\langle{\mathcal{T}}\rangle$ and
$Q_0^{\mathcal{T}}=1$. Plugging the ansatz (\ref{hp}) on the
kinematic residue equations (\ref{3}) and (\ref{kre}) produces a
set of recursive equations for the constants
$H_{k}^{\mathcal{T}|1\ldots 1}$ and the functions
$Q_{k}^{\mathcal{T}|1\ldots 1}(x_1, \ldots, x_k)$ which may be
solved for specific models.

So far, the only model for which all
solutions to these equations are known is the Ising field theory.
The solutions were obtained in \cite{nexttonext}. Since we are
dealing the free Fermion case, it is natural to expect that the
form factors of the twist field would admit closed expressions in
terms of Pfaffians, as for the order and disorder fields of the
Ising theory. This is indeed the case, and it is easy to show that
\begin{equation}\label{f}
F_{k}^{\mathcal{T}|11\ldots 1}(\theta_1,
\ldots,\theta_{k})=\langle\mathcal{T}\rangle{\rm Pf}(K),
\end{equation}
where ${\rm Pf}$ is the Pfaffian, which is mainly characterised by
the property that ${\rm Pf}(K)^2 = \det(K)$ and $K$ is an
anti-symmetric $k \times k$ matrix, with entries
\begin{equation}\label{k}
  K_{ij}=K(\theta_{ij}).
\end{equation}
Notice that, for the Ising model only $k$-particle form factors
with $k$ even are non-vanishing. The function $K(x)$  has
properties
\begin{eqnarray}
  K(\theta) &=& -K(-\theta), \label{p1}\\
  K(\theta)|_{n=1} &=& 0, \label{p2}\\
  \lt(K(\theta+ i s)\rt)^* &=& -K(\theta-is), \qquad \theta, s \in
  \mathbb{R},\label{p3}
\end{eqnarray}
where ``*" indicates complex conjugation. The Pfaffian expression
is nothing else than the application of Wick's theorem on the
particles in the asymptotic states (specialised to all particles being on copy 1), a contraction of
two particles being $K(\theta_{12})$.

\subsection{Form factor consistency checks: the $\Delta$-sum
rule}

The form factor program above provides form factor solutions,  but
it is not obvious that these solutions are the correct ones for
the branch-point twist fields $\tw$ and $\t\tw$. Indeed, as we
mentioned, the twist property and invariance under other
symmetries alone are not enough to fully fix the field; we need to
specify its scaling dimension to be the minimal one. It is
generally conjectured that the ``minimal'' solution corresponds to
a primary field (hence with minimal scaling dimension in a given
Virasoro module), but it is always good to verify that the
solutions found agree with the expected scaling dimension obtained
in the ultraviolet limit. As is well-known, the form factor
program provides a way of carrying out this verification by
allowing us to compute the correlation functions of various fields
of an integrable QFT. In the ultraviolet limit, it is possible to
relate a particular correlation function to the holomorphic
conformal dimension $\Delta$ (with $\Delta+\b\Delta = d$, the
scaling dimension, and $\Delta-\b\Delta = s$, the spin) of a
primary field by means of the so-called $\Delta$-sum rule:
\begin{equation}\label{delta}
    \Delta^{\tw} = \Delta^{\t\tw}=-\frac{1}{2\langle \mathcal{T}
    \rangle}\int_{0}^{\infty} r
    \left\langle \Theta(r) \t\tw(0)  \right\rangle dr
\end{equation}
(where the integration is on a space-like ray), originally
proposed by G.~Delfino, P.~Simonetti and J.L.~Cardy in \cite{DSC}. Here,
where $\Theta$ is the local operator corresponding to the trace of
the stress-energy tensor. The first equality, naturally expected from CFT,
holds from the $\Delta$-sum rule thanks to the fact that $\Theta$
commute with $\tw$ and that $\Theta^\dag=\Theta$. In the cases of the branch-point twist fields, which are spinless, the holomorphic
conformal dimension is related to the scaling dimension by $d_n =
2\Delta^\tw$, where $d_n$ is expected to be (\ref{scdim}).

Introducing the complete sum over quantum states (\ref{csum}) and carrying out the $r$-integration, the expression above can be
rewritten as
\begin{eqnarray}
\Delta^{\mathcal{T}} &=&-\frac{1}{%
2\left\langle \mathcal{T}\right\rangle }\sum_{k=1}^{\infty
}\sum_{\mu _{1}\ldots \mu _{k}}\int\limits_{-\infty }^{\infty
}\ldots
\int\limits_{-\infty }^{\infty }\frac{d\theta _{1}\ldots d\theta _{k}}{%
k!(2\pi )^{k}\left( \sum_{i=1}^{k}m_{\mu _{i}}\cosh \theta
_{i}\right) ^{2}}
\nonumber \\
&&\times F_{k}^{\Theta |\mu _{1}\ldots \mu _{k}}(\theta
_{1},\ldots ,\theta _{k})\,\left( F_{k}^{\mathcal{T}|\mu
_{1}\ldots \mu _{k}}(\theta _{1},\ldots ,\theta _{k})\,\right)
^{*}\,\,\,, \label{dcorr}
\end{eqnarray}
where the sum in $\mu_i$ with $i=1,\ldots, k$ is a sum over
particle types in the theory under consideration. The sum starts
at $k=1$ since we are considering ``connected" correlation
functions, that is, the $k=0$ contribution has been subtracted.
The sum above, can only be carried out in particularly simple
cases. For most models, one must be content with evaluating just
the first few contributions to the sum. Fortunately, the many
studies carried out in the last years provide strong evidence that
the sum above is convergent and that in fact, the first few terms
provide the main contribution to the final result. Indeed, the
convergence is often so good that considering only the
contribution with $k=2$ already provides very precise results (see
e.g. \cite{FMS}). Expecting a similar behaviour also in our case,
we will approximate the sum above by the two-particle
contribution, that is
\begin{eqnarray}
\Delta^{\mathcal{T}}  \approx -\frac{n}{ 2\left\langle
\mathcal{T}\right\rangle } \int\limits_{-\infty }^{\infty
}\int\limits_{-\infty }^{\infty }\frac{d\theta _{1} d\theta _{2}
F_{2}^{\Theta |11}(\theta _{12}) F_{2}^{\mathcal{T}|11}(\theta
_{12})^{*}}{2 (2\pi )^{2} m^2 \left( \cosh \theta _{1} + \cosh
\theta_2 \right) ^{2}}
 . \label{dcorr2}
\end{eqnarray}
The factor of $n$ is a consequence of summing over all particle
types and using (\ref{use}). In addition, the only non-vanishing
contribution comes from form factors involving only one particle
type, since we are considering $n$ non-interacting copies of the
model. This implies that
\begin{equation}
   F_{2}^{\Theta|ij}(\theta)=0\qquad \forall \qquad i\neq j.
\end{equation}
Changing variables to $\theta=\theta_1-\theta_2$ and
$\theta'=\theta_1+\theta_2$ we obtain,
\begin{equation}\label{DeltaInt}
\Delta^{\mathcal{T}} \approx -\frac{n}{32 \pi^2 m^2 \left\langle \mathcal{T}\right\rangle }
\int\limits_{-\infty}^{\infty }d \theta\, \frac{F_{2}^{\Theta
|11}(\theta) F_{2}^{\mathcal{T}|11}(\theta)^{*}}{\cosh^2(
\theta/2)}.
\end{equation}
In \cite{entropy} we evaluated the integral above both for the
Ising and sinh-Gordon models, whereas in \cite{other} we performed
a similar computation for the sine-Gordon model. We briefly
summarise below our results for the Ising and sinh-Gordon
theories.

\subsubsection{The $\Delta$-sum rule for the Ising and sinh-Gordon models}

 For the Ising model the only non-vanishing form
factor of the trace of the stress-energy tensor is the
two-particle form factor. Hence, the two-particle approximation
(\ref{dcorr2}) becomes exact. The two-particle form factors are
given by
\begin{equation}
F_{2}^{\mathcal{T}|11}(\theta)=\frac{-i  \langle
\mathcal{T}\rangle \cos\left(\frac{\pi}{2n}\right)}{ n
\sinh\left(\frac{ i\pi +
\theta}{2n}\right)\sinh\left(\frac{i\pi-\theta}{2n}\right)}
{\sinh\left(\frac{\theta}{2n}\right)},\qquad
F_{2}^{{\Theta}|11}(\theta)= -2 \pi i m^2
\sinh\left(\frac{\theta}{2}\right),
\end{equation}
and therefore
\begin{eqnarray}
\Delta^{\mathcal{T}} = -\frac{1}{16 \pi} \int\limits_{-\infty
}^{\infty }\frac{
\cos\left(\frac{\pi}{2n}\right)\sinh\left(\frac{\theta}{2n}\right)
\sinh\left(\frac{\theta}{2}\right)}{ \sinh\left(\frac{ i\pi +
\theta}{2n}\right)\sinh\left(\frac{i\pi-\theta}{2n}\right)\cosh^2\left(\frac{\theta}{2}\right)}\,
d \theta .
\end{eqnarray}
The above integral can be computed analytically for $n$ even by
shifting $t$ by $2 \pi n i$ and noticing that the integral changes
by a sign, so that
\begin{equation}\label{quiteknut}
    2\Delta^{\mathcal{T}}=2 \pi i\sum_{i=1}^n r_j=\frac{1}{24}\left(n-\frac{1}{n}\right),
\end{equation}
where $r_j$ are the residues of the poles of the integrand at $t=i
\pi (2j-1)$, with $j=1,\ldots,n$.  These residues can be easily
computed (see \cite{entropy} for the details) and the expected
result (\ref{scdim}) is reproduced, with $c=1/2$.

For the sinh-Gordon model, the relevant $2$-particle
form factors are given by
\begin{equation}
F_{2}^{\mathcal{T}|11}(\theta)=\frac{ \langle \mathcal{T}\rangle
\sinh\left(\frac{\pi}{n}\right)}{2 n \sinh\left(\frac{ i\pi +
\theta}{2n}\right)\sinh\left(\frac{i\pi-\theta}{2n}\right)}
\frac{F_{\text{min}}^{\mathcal{T}|11}(\theta)}{F_{\text{min}}^{\mathcal{T}|11}(i \pi)},\qquad
F_{2}^{{\Theta}|11}(\theta)= 2 \pi m^2
\lt. \frac{F_{\text{min}}^{\mathcal{T}|11}(\theta)}{F_{\text{min}}^{\mathcal{T}|11}(i \pi)}\rt|_{n=1}.
\end{equation}
The form factors of $\Theta$ were computed in \cite{FMS}. Since
$\Theta$ is a local operator, its minimal form factor is given by
(\ref{int}) with $n=1$.
The tables below show the result of carrying out the integral (\ref{DeltaInt})
numerically for various values of $n$ and $B$. Next to each value
of $n$ in brackets we show for reference the expected value of
$\Delta^{\mathcal{T}}$, as predicted by the CFT formula
(\ref{scdim}) (with, again, $\Delta^\tw = d_n/2$).
\begin{center}
\begin{tabular}{|l|l|l|l|l|}
\hline & $n=2$ (0.0625) & $n=3$ (0.1111) & $n=4$ (0.1563) & $n=5$ (0.2)\\
\hline
 $B=0.02$& 0.0620& 0.1114& 0.1567  & 0.2007 \\ \hline
 $B=0.2$ & 0.0636 & 0.1135& 0.1599  & 0.2048  \\\hline
 $B=0.4$ & 0.0636& 0.1148 & 0.1620   & 0.2074   \\\hline
 $B=0.6$ & 0.0643&0.1155 & 0.1631   &0.2088   \\\hline
 $B=0.8$ & 0.0644&0.1158 & 0.1636  & 0.2096 \\\hline
 $B=1$  & 0.0644& 0.1159 & 0.1637 &0.2098   \\\hline
\end{tabular}
\end{center}
\begin{center}
\begin{tabular}{|l|l|l|l|l|l|}
\hline & $n=6$ (0.2431)& $n=7$ (0.2857)& $n=8$ (0.3281)& $n=9$ (0.3704) & $n=10$ (0.4125)\\
\hline
 $B=0.02$ & 0.2436   &  0.2864 & 0.3289 & 0.3712  &0.4135  \\ \hline
 $B=0.2$ & 0.2488  & 0.2925  &0.3360   & 0.3793  & 0.4225 \\\hline
 $B=0.4$ & 0.2522  &0.2966   & 0.3407  & 0.3846 & 0.4284 \\\hline
 $B=0.6$ & 0.2540 & 0.2988  &  0.3433 & 0.3876  & 0.4317 \\\hline
 $B=0.8$ &0.2550  & 0.2999  & 0.3446 & 0.3890 & 0.4334 \\\hline
 $B=1 $ & 0.2552  & 0.3002   & 0.3449  & 0.3895 & 0.4339\\\hline
\end{tabular}
\end{center}
The figures obtained are extremely close to their expected value
for all choices of $B$ and $n$.

\sect{Bulk and boundary entropy from twist field form factors}\label{sectent}

Formulas (\ref{rhon}) and (\ref{be3}) express the entanglement
entropy in  the bulk and boundary cases, respectively, in terms of
the two-point and one-point functions of branch-point twist
fields. The form factor expansions (\ref{bulkexp}) and
(\ref{boundexp}) then express these correlation functions as
infinite series obtained from form factors (\ref{ff}). These
infinite series provide exact expressions, but also, as we
explained in subsection \ref{ssectffexp}, the truncated, partial
series give efficient large-distance expansions. Our main result,
(\ref{main}), is an exact large-distance correction for the bulk
entanglement entropy, and is obtained by keeping, in the series
(\ref{bulkexp}), only the two-particle contributions. On the other
hand, our analysis of the boundary entanglement entropy, in the
Ising model, required the consideration of all terms in
(\ref{boundexp}). We will now explain how to go from form-factor
expansions to entanglement entropy.

\subsection{Bulk entanglement entropy}

From expression (\ref{bulkexp}) and from (\ref{rhon}), we find for the bulk entanglement entropy:
\beq\label{bulkfinal}
    S_A^{\rm bulk}(rm) = -\frac{c}{3} \log(\varep m) + U^{\text{model}} + \sum_{k=1}^\infty
    e_k(rm),
\eeq where $U^{\text{model}}$ was defined in (\ref{Ugen}), and
\begin{equation}\label{elgen}
    e_k(rm)=-\lim_{n\rightarrow 1} \frac{d}{dn} \left[\sum_{k=0}^{\infty}\sum_{\mu_1,\ldots,\mu_{k}=1}^n
\int\limits_{\theta_1>\cdots>\theta_k} \left[\prod_{j=1}^{k}\frac{d\theta
_{j}}{ 2\pi}\,e^{-rm_{\mu_j}
\cosh\theta_{j}}\right]\left|F_{k}^{\mathcal{T} |\mu_1 \mu_2
\ldots \mu_{k}}(\theta _{1},\ldots,\theta_{k})\right|^2\right].
\end{equation}
The expression for $e_k$ could be re-written in various ways, by
performing the sum in one of particle indices $\mu_i$ using
invariance under shift of copy numbers, and/or by using the
formula (\ref{tralala}) in integrable models. Also, again in
integrable models, the range of rapidity integration can easily be
extended to $-\infty<\theta_j<\infty,\,j=1,2,\ldots,k$ (i.e.\
without ordering of rapidities), putting a factor $1/k!$. Indeed,
the exchange of two rapidities in the form factors just
corresponds to a multiplication by a two-particle $S$-matrix,
according to (\ref{1}). By unitarity,
$|S_{\mu_1\mu_2}(\theta)|^2=1$, so that the integrands in
(\ref{bulkexp}) are the same under exchange of rapidities. Another
way of understanding this is that with different ordering of
rapidities, we simply have a different basis of states: instead of
in-states, we have states where in- and out-configurations are
mixed. We may also sum over these states in order to get the
form-factor expansion, or we may average over many such bases,
integrating in the full rapidity range. Such mixed bases exist as
asymptotic-state bases in integrable models thanks to the
independence of the scattering amplitudes upon impact parameters
(see e.g. \cite{Mussardo:1992uc}). In non-integrable models, mixed
bases are not expected to exist. However, thanks to
(\ref{ni1gen}), the latter argument can still be used, for the
two-particle contribution, to extend the integration region to the
whole rapidity range, without ordering. This will be useful in the
next paragraph, where we study in more details the two-particle
contribution.

\subsubsection{Next-to-leading order IR correction to the bulk entanglement entropy}

From the expression above (\ref{bulkfinal}), it appears that the
first correction to the IR behaviour of the entanglement entropy
would come from the one-particle form factor contribution
\begin{equation}\label{e1}
     e_1(rm)=-\lim_{n\rightarrow 1} \frac{d}{dn}\left[n
\int\limits_{-\infty }^{\infty } \frac{d\theta}{2\pi}\,e^{-rm
\cosh\theta} \left|F_{1}^{\mathcal{T}
|1}(\theta)\right|^2\right].
\end{equation}
For all theories studied in \cite{entropy,other} this correction
vanishes because their internal symmetries are such that only form
factors of even particle numbers are non-vanishing. However, it is
possible to argue that even when the one-particle form factor is
not zero, its  contribution to the entropy should be. First, the
one-particle form factor is always independent of the rapidity due
to relativistic invariance and the spinlessness of the twist
field. Hence, it only gives a simple $n$-dependent factor. Second,
the form factor is zero at $n=1$, as in that case the twist field
is simply the identity. If we assume that it goes to zero like
$n-1$ (or in fact, faster than $\sqrt{n-1}$), then we can conclude
that the derivative of its square at $n=1$ must also vanish.
Certainly, however, it would be desirable to evaluate exactly
one-particle form factors in models where they do not vanish by
symmetry argument, in order to verify this.

We will now look at the next correction (the first non-trivial
one), namely the correction associated to the two-particle form
factor contribution
\begin{eqnarray}
e_2(rm) &=& -\lim_{n\rightarrow
1}\frac{d}{dn}\left[n\sum_{j=0}^{n-1}\int\limits_{-\infty
}^{\infty } \int\limits_{-\infty }^{\infty }\frac{d\theta
_{1}d\theta _{2}}{ 2!(2\pi )^{2}}\left|F_{2}^{\mathcal{T}
|11}(\theta_{12}+2 \pi i j)\right|^2\,e^{-rm
(\cosh\theta_1 + \cosh \theta_2)}\right]\nonumber\\
&=&-\lim_{n\rightarrow
1}\frac{d}{dn}\left[\langle\mathcal{T}\rangle^2\frac{n}{4
\pi^2}\int\limits_{-\infty }^{\infty } d\theta f(\theta,n)
K_{0}(2rm\cosh(\theta/2))\right]..\label{ent1}
\end{eqnarray}
In the first line, we extended the rapidity integration range so that there is no ordering
of rapidities (dividing by $2!$) as described above,
we summed over one particle index using invariance under shift
of copy numbers, and we used
formula (\ref{use2}). Notice that these operations are valid
both at and out of integrability.
In the second line we just carried out
one of the integrals, where $K_0(x)$ is a modified Bessel function
of the second kind, and we define:
\begin{equation}
    f(\theta,n)=\langle\mathcal{T}\rangle^{-2}\left(\left|F_{2}^{\mathcal{T}
    |11}(\theta)\right|^2+ \sum_{j=1}^{n-1}\left|F_{2}^{\mathcal{T} |11}(\theta+2 \pi i
j)\right|^2 \right).\label{ftn}
\end{equation}
For simplicity, we assumed that there is only one particle type,
so that both particles involved in this two-particle
contribution have the same mass. As we will see, this turns out to be
enough in order to obtain the general result.

What follows will be devoted to show that, if we
denote by  $\t{f}(\theta,n)$ the analytic continuation of
$f(\theta,n)$ to non-integer values of $n$, then: \beq
   -\lim_{n\rightarrow 1}\frac{d}{dn} \left[n\t{f}(\theta,n)
   \right]=-\frac{\pi^2\delta(\theta)}{2},
\eeq
   which implies
\begin{equation}
   e_2(rm) =-\frac{K_0(2mr)}{8}.\label{rs}
\end{equation}
Recall that $f(\theta,1)=0$ and that $f(\theta,n)$ is only defined
for integer values of $n$. It is also easy to see, from the
general solution for the two-particle form factors (\ref{full}) or
from the general expectation that the form factors vanish like
$n-1$ as $n\to1$, that the term $|F_{2}^{\mathcal{T}
|11}(\theta)|^2$ will not contribute to the derivative at $n=1$.

In order to obtain the entropy, we should now analytically
continue the expression inside the derivative in (\ref{ent1}), as function of $rm$
and $n$, from $n=1,2,3,\ldots$ to $n\in[1,\infty)$, compute the
derivative with respect to $n$ and evaluate the result at $n=1$.
The analytic continuation is of course not unique. The motivation for the choice of analytic continuation that was used in \cite{entropy,other} was based on Carlson's theorem, and on an expectation about the large-$n$ behaviour of partition functions on multi-sheeted Riemann surfaces (or spaces with conical singularities). Essentially, we want the analytic continuation of $f(\theta,n)$, for any fixed $\theta$, to be well-behaved enough at large $n$. In the integrable cases studied with precision in \cite{entropy} and \cite{other}, we found that there was an analytic continuation such that the divergence at large $n$ is less than exponential for any ${\rm Re}(n)>0$. This is a unique analytic continuation with this property by Carlson's theorem.

\begin{figure}[h!]
\bc \includegraphics[width=11cm,height=7.5cm,angle=0]{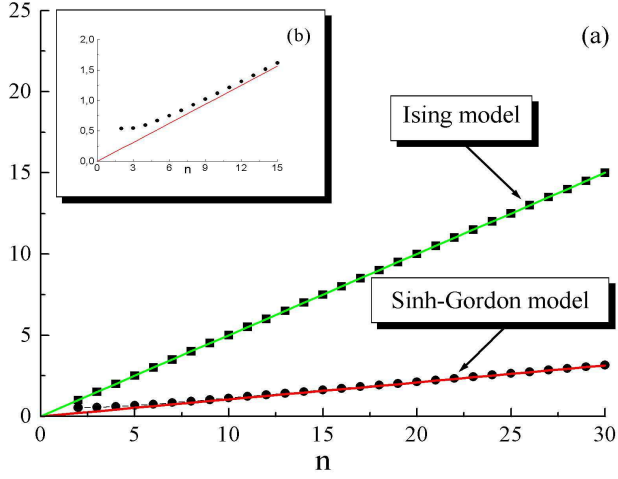}\ec
    \caption{(a) shows 4 functions: the points are the function $n f(0,n)$ for integer
values of $n$ in the interval $[2,30]$ both for the Ising (black
squares) and sinh-Gordon (black circles) models, evaluated
numerically. The solid green line gives the analytic continuation
$n \tilde{f}(n)$ for real values of $n$ in the interval $[0,30]$
for the Ising model, that is the function $n/2$. Notice that $
\tilde{f}(1)=1/2$ whereas $f(0,1)=0$. Finally, the solid red line
gives the function $n \tilde{f}(\infty)$ for the sinh-Gordon
model, that is a straight line passing through the origin which
describes the asymptotic behaviour of the function $n
\tilde{f}(n)$ for $n$ large. In the sinh-Gordon case, all
functions have been computed for $B=0.5$. (b) is a magnification
of the lower left corner of the sinh-Gordon part of (a) which
demonstrates (\ref{ident}), namely, as for the Ising model,
$\tilde{f}(1)=1/2$ and $f(0,1)=0$. For the sinh-Gordon model, the
analytic continuation is $n \tilde{f}(n)=a_0 n + \sum_{i=1}^\infty
\frac{a_{i}}{n^{2i-1}}$, for some coefficients $a_0, a_1, \ldots$
which depend of the coupling $B$ \cite{entropy}. Such analytic
continuation was also studied for the sine-Gordon model in
\cite{other} and found to display a power series structure, with
powers depending on the values of the coupling constant, very
different from the one identified for the sinh-Gordon model.}
\label{fig-fn}
\end{figure}

Once we admit this requirement on the analytic continuation,
the evaluation of the derivative at $n=1$ can be done exactly.
The observation at the basis of this evaluation is that although
the analytic continuation of the sum in (\ref{ftn})
vanishes as $n\to1$ for all $\theta\neq0$, it does not vanish as $n\to1$ when $\theta=0$.
That is, the function $\t{f}(\theta,n)$ does not convergence uniformly in $\theta$ as $n\to1$.
Let us define for convenience
\beq
    \t{f}(n):=\t{f}(0,n).
\eeq
What we find is that its limit $n\to 1$
is non-zero, and positive. Note that this means that in fact, $f(0,n)$ does not have a
well-behaved analytic continuation from $n=1,2,3,\ldots$ to $n\in[1,\infty)$.
Rather, $\t{f}(n)$ is such an analytic continuation, but
from $n=2,3,\ldots$ to $n\in[1,\infty$; it has the property that
$\t{f}(1)\neq f(0,1)=0$.
This was observed explicitly in the Ising and sinh-Gordon model \cite{entropy}, where we found $\t{f}(n) = 1/2$
in the Ising model, and obtained a large-$n$ expansion for $\t{f}(n)$ in the sinh-Gordon model from
which we could evaluate with some precision its value at $n=1$.
It was also observed in the sine-Gordon model \cite{other},
using this time a power series in $n$. See Fig. \ref{fig-fn}.

Naturally, the fact that the function $\t{f}(\theta,n)$ does not
converge to 0 as $n\to1$ at the point $\theta=0$
is of no importance if we just want to evaluate the integral
over $\theta$ (with, as factor, a function that is
well-behaved at $\theta=0$). Indeed, the point $\theta=0$ is of
Lebesgue measure zero, so the value of the non-uniform limit
at that point, if it is finite, does not affect the integral. However, we want to
take the derivative with respect to $n$.
It is possible to argue that this derivative $\frc{\p}{\p
n}\t{f}(\theta,n)$ at $n=1$ should actually be proportional to a
{\em delta-function} $\delta(\theta)$: it has non-zero support at $\theta=0$.
Intuitively, this is because looking at the value of $\t{f}(\theta,n)$
near enough to $\theta=0$, we see that it should vary very fast as $n\to1$, since
it should go to zero for any $\theta\neq0$, it should stay finite for $\theta=0$, and it should
be continuous in $\theta$ for all $n>1$. This very fast variation becomes an
infinite variation as $n\to1$ for $\theta$ infinitesimally near to 0, which
produces a $\delta(\theta)$. Note that the resulting derivative at $n=1$ should also be positive, since this fast variation
goes from $\t{f}(0)>0$ to 0 as $n$ decreases to 1. This gives
a negative correction to the saturation of the entanglement entropy, which is
indeed expected on physical grounds.

\begin{figure}
\bc
\includegraphics[width=14cm,height=4cm]{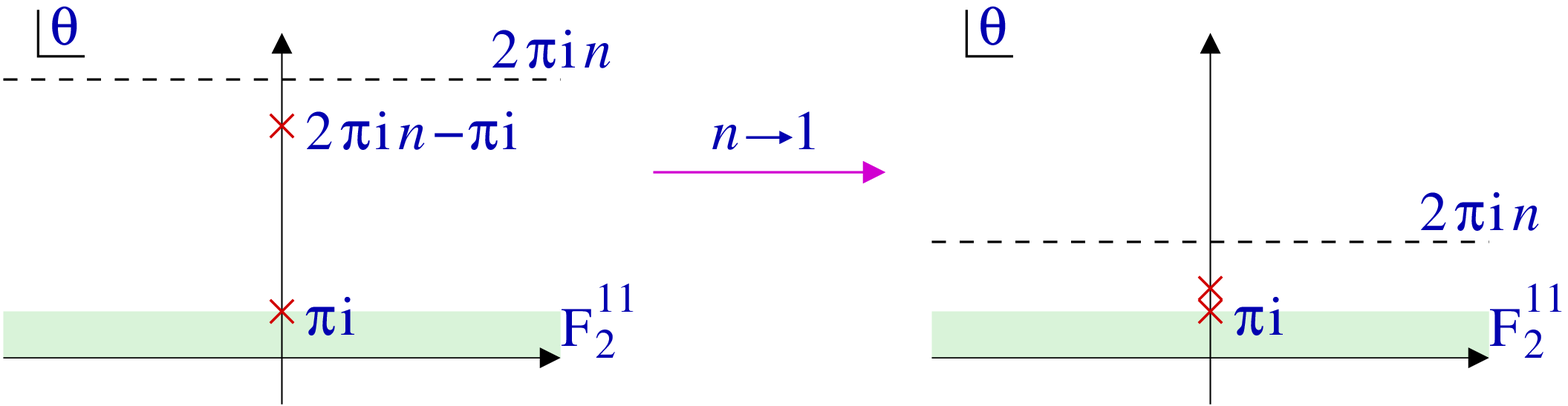}
\ec \caption{The collision of kinematic singularities when
$n\to1$.} \label{fig-collide}
\end{figure}

But why would the function $\t{f}(\theta,n)$, as function of $\theta$, not
converge uniformly as $n\to 1$? In order to answer this, we have to find a way
of analytically continuing in $n$ the summation term in (\ref{ftn}). There are a great
many ways of doing this, and in all cases, what we find is that the cause for the non-uniform
convergence is the collision of kinematic poles, at $\theta=i\pi$ and $\theta=2i\pi n-i\pi$,
as $n\to1$, see Fig. \ref{fig-collide}.
For instance, using a Poisson resummation formula to transform the
sum over $j$ into an integral \cite{entropy}, one sees that the
kinematic poles pinch this integral as $n\to1$. Another way is to
perform the summation via a contour integral with a cotangent
kernel whose residues reproduce the terms that are to be summed.
There, one has to extract the poles of the form factors
themselves, these extracted poles collide as $n\to1$. Both ways
necessitate periodicity of the function $\t{f}(\theta,n)$ upon
adding $2i\pi n$ to $\theta$. The latter way was in fact
generalised to non-integrable models in \cite{next}.

Yet, perhaps the simplest way of extracting the derivative is simply to
extract the kinematic poles themselves, and then to perform the summation
exactly on these poles \cite{entropy}.
The contribution of the kinematic singularities to the sum in the
function $\t{f}(\theta,n)$ (\ref{ftn}) is obtained from the
singular behavior in $j$ of the summand $s(\theta,j)=
F_{2}^{\mathcal{T} |11}(\theta+2\pi i j) \lt(F_{2}^{\mathcal{T}
|11}\rt)^*(\theta-2\pi i j)$: \[
    s(\theta,j)\sim \frc{i\;F_{2}^{\mathcal{T} |11}(2\theta+2\pi i n-i\pi)}{\theta-2\pi i j+2\pi i n-i\pi}
        - \frc{i\;F_{2}^{\mathcal{T} |11}(2\theta+i\pi)}{\theta-2\pi i j+i\pi} + {\rm c.c.}
\]
(where ${\rm c.c.}$ means complex conjugate, taken for real $\theta$ and real $j$).
It is a simple matter to perform on this expression the sum
$\sum_{j=1}^{n-1}$, giving \beqa
    \sum_{j=1}^{n-1} s(\theta,j) &\sim& \frc1{2\pi} \lt(\psi\lt(-\frc12+n-\frc{i\theta}{2\pi}\rt) - \psi\lt(\frc12-\frc{i\theta}{2\pi}\rt)\rt)
        F_{2}^{\mathcal{T} |11}(2\theta+2\pi i n-i\pi) \n &&
        + \frc1{2\pi} \lt(\psi\lt(-\frc12+n+\frc{i\theta}{2\pi}\rt) - \psi\lt(\frc12+\frc{i\theta}{2\pi}\rt)\rt)
        F_{2}^{\mathcal{T} |11}(2\theta+i\pi) \n && + {\rm c.c.} \no
\eeqa where $\psi(z) = d\log \Gamma(z) /dz$ is the derivative  of
the logarithm of Euler's Gamma function. This has no poles at
$\theta=0$, as the kinematic poles of the form factors involved that would produce a pole at $\theta=0$ actually
cancel out. The poles that are nearest to ${\rm Re}(\theta)=0$ as
$n\to1$ are at $\theta = \pm i\pi(n-1)$, again coming from the form
factors involved. The residues to first order in $n-1$ give:
\beq\label{tfnm}
    \t{f}(\theta,n) \sim \t{f}(1) \lt( \frc{i\pi(n-1) }{2(\theta+i\pi(n-1))} - \frc{i\pi(n-1) }{2(\theta-i\pi(n-1))}\rt) \quad (n\to1)
\eeq with \beq
    \t{f}(1) = \frc12~.
\eeq This has simple poles at $\theta=\pm i\pi(n-1)$ with residues
that vanish at $n=1$, gives $\lim_{n\to1} \t{f}(0,n) = \t{f}(1)$, and vanishes
like $(n-1)^2$ as $n\to1$ for $\theta\neq0$. The limit $n\to1$ of the derivative with
respect to $n$, as
a distribution on $\theta$, is easily evaluated: \beq
    \lt(\frc{\p}{\p n} \t{f}(\theta,n)\rt)_{n=1} = \pi^2 \t{f}(1)
    \delta(\theta)~.\label{ident}
\eeq The full form of $\t{f}(\theta,n)$ was obtained in appendix C
of \cite{entropy}. For the free case our result is in agreement
with the $n \rightarrow 1$ limit evaluated previously in
\cite{casini1}.

Inserting this inside (\ref{bulkfinal}) we find that the
entanglement entropy $S_A(rm)$ for $A$ an interval of length $r$
is:
\begin{equation}
    S_A^{\text{bulk}}(rm)=-\frac{c}{3}\log(\varepsilon m) + U^{\text{model}} - \frc{1}{8} K_0(2rm) +
    O\lt(e^{-3rm}\rt),
\end{equation}
with $ U^{\text{model}}$ defined in (\ref{Ugen}). This result can
be generalised to theories with more than one particle and/or
bound states \cite{other}: the bound states poles never collide,
hence do not provide additional contributions; and the kinematic
poles only occur for particle-antiparticle form factors, so that
we have a contribution for all particle types of the theory. The
result was also generalised to non-integrable models: the only
properties needed to derive the result were those seen in
subsection \ref{ssectni} to hold out of integrability. Hence, we
obtain the more general result (\ref{main}) quoted in the
introduction.

In general, the constant $U^{\text{model}}$ is extremely hard to
evaluate, even in integrable models. However, for models of free
particles like the Ising model, it can be evaluated by standard
methods of angular quantisation \cite{Luky1,Luky2}. This was done
in \cite{entropy} for the Ising model (by simply relating the
branch-point twist fields to ordinary $U(1)$-twist fields for
which the calculation had already been done), where we obtained
the value (\ref{u}).

A final note is in order concerning the function $\t{f}(n)$. Above we concentrated on its
behaviour as $n\to1$, as this is what matters for the evaluation of the entanglement entropy.
However, the behaviour of this function as $n\to\infty$ is also somewhat interesting.
Indeed, in all models studied where the perturbation from the UV fixed point was strictly relevant (Ising, sinh-Gordon
and sine-Gordon models), the behaviour at large $n$ was observed to be exactly linear. However,
for the sine-Gordon model at a value of the coupling that makes the perturbation marginally
relevant, we observed a behaviour like $n\log n$ \cite{other}. Interestingly, then, it seems that the
large-$n$ behaviour encodes information about the short-distance theory: the way
the model behaves near to the critical point.

\subsubsection{Higher order IR corrections to the bulk entanglement entropy}

In order to evaluate higher order corrections to the entropy,
namely $e_3(mr), e_4(mr) \ldots$  we need first of all to know
the form factors for more than two particles. As indicated
in previous sections, such solutions are only known for the Ising
model (see (\ref{f})) for which all higher particle corrections to
the entropy were evaluated in \cite{nexttonext}. Once more, the
problem of finding the right analytic continuation in $n$ plays a
fundamental role. Solving this problem has been indeed, even for
the Ising model, extremely challenging (see \cite{nexttonext} for
the details of this). Although we have not investigated
interacting models yet, we suspect that higher order corrections
may also be, at least to some extent, of a universal nature, as
the pole structure of the form factors plays again a key role. For
the Ising model we have found that,
\begin{eqnarray}\label{el}
 e_{2k}(rm) &=&
    \frac{ \pi^2 (-1)^{k} }{
    k}\left[\prod_{a=1}^{2k} \int_{-\infty}^{\infty} \frac{d \theta_{a}}{4\pi}\rt]\delta(\theta)
     \lt[
\left(%
\begin{array}{c}
 2k-2 \\
  k-1 \\
\end{array}%
\right) \prod\limits_{j=1}^{2k}  \frc{e^{-rm \cosh \theta_{j}} }{
    \cosh\frac{\hat{\theta}_{j j+1}}{2}}\right. \nonumber \\ &&  \left. -
 \sum_{j=1}^{k}\sum_{a=1}^{j-1} \sum_{q=\pm}
 \left(%
\begin{array}{c}
  2k-1 \\
  k-j \\
\end{array}%
\right)     (-1)^{j} \prod\limits_{i=1}^{2k}  \frc{e^{-rm \cosh
(\theta_{i}+ q\frc{j-a}{2k}\pi i)}
 }{\cosh\lt(\frac{\h\theta_{i,i+1}}{2} + q\frc{j-a}{2k} \pi i\rt)}
 \rt],
\end{eqnarray}
and $e_{2k +1}(rm)=0$, where $\theta=\sum_{i=1}^{2k} \theta_i$ and
$\hat{\theta}_{ij}=\theta_i+\theta_j$. Notice that for $k=1$ we
exactly recover the result (\ref{rs}).

\subsection{Boundary entanglement entropy}

In order to make use of the form factor expansion (\ref{boundexp}) in the entanglement
entropy (\ref{be3}), we need to describe in more detail the boundary state
introduced in section \ref{bent}. As mentioned there, the state
$|B\rangle$ is just a tensor product of boundary states in the
individual copies. In integrable models, these have an explicit
expression as the famous boundary state introduced by Ghoshal and
Zamolodchikov \cite{Ghoshal:1993tm}. In the case where no boundary
bound state can form, we have
\begin{equation}\label{bs}
|B\rangle =\exp\left(\frac{1}{4
\pi}\sum_{\mu=1}^{n}\int_{-\infty}^{\infty}
R\left(\frac{i\pi}{2}-\theta\right) Z_{\mu}(-\theta)Z_\mu
(\theta)\right)|0\rangle.
\end{equation}
The function $R(\theta)$ is the boundary reflection matrix of the
integrable QFT (assuming still it has just one particle) and
$Z_\mu(\theta)$ are the Faddeev-Zamolodchikov operators, which
provide a generalization of the creation-annihilation operators
for integrable QFTs with non-trivial interactions
\cite{ZZ,Faddeev:1980zy}. Their main properties are \beqa
    && Z_{\mu_1}(\theta_1) \cdots Z_{\mu_k}(\theta_k) |0\rangle = |\theta_1,\ldots,\theta_k\rangle_{\mu_1,\ldots,\mu_k} \mbox{ for }
    \theta_1>\ldots>\theta_k \n
&&Z_{\mu_1}(\theta_1)Z_{\mu_2}(\theta_2) = S_{\mu_1\mu_2}(\theta_1-\theta_2)
Z_{\mu_2}(\theta_2)Z_{\mu_1}(\theta_1). \no \eeqa The tensor-product form of
the boundary state indicates that particles living in different
copies of the theory do not interact through the presence of the
boundary.

Using the boundary operator defined above, we can evaluate all boundary-state overlaps
occurring in (\ref{boundexp}) and obtain a large-$r$ expansion for the entanglement entropy
from (\ref{be3}) in terms of form factors, as for the bulk case. In fact, we could see this
expansion simply as coming from expanding in (\ref{be3}) the exponential defining the boundary state
in (\ref{bs}). Either ways, this gives, with the distance between $x$ and the boundary being $r$,
\begin{equation}\label{sk}
 \langle 0| {\mathcal{T}}(x)|B \rangle = \bra{\cal T}\ket \sum_{k=0}^{\infty}
    f_{2k}(2rm),
\end{equation}
where
\begin{eqnarray}\label{1p}
    \bra{\cal T}\ket f_{2k}(t)&=&\frac{1}{k!
    (4\pi)^k}\sum_{j_1,j_2,\ldots,j_k=1}^{n} \left[\prod_{r=1}^{k} \int_{-\infty}^{\infty} d \theta_{r}e^{-t \cosh \theta_{r}}
    R\left(\frac{i\pi}{2}-\theta_{r}\right)\right]\nonumber
    \\
    & \times & F_{2k}^{\mathcal{T}|j_1 j_1 j_2 j_2\ldots j_k j_k}
    (-\theta_{1},\theta_{1},\ldots,-\theta_{k},\theta_{k}).
\end{eqnarray}
For example:
\begin{eqnarray}
  f_0(t)
    &=&  1, \\
  f_2(t) &=&\frac{n}{4
    \pi \bra{\cal T}\ket}\int_{-\infty}^{\infty}d\theta\,
    R\left(\frac{i\pi}{2}-\theta\right)
    F_{2}^{\mathcal{T}|11}(-2\theta)e^{-t \cosh \theta} ,\label{2p}
\end{eqnarray}
and so on. Therefore, we can write the boundary entanglement
entropy as
\beq\label{boundaryfinal}
    S_A^{\rm boundary}(rm) = -\frac{c}{6} \log(\varepsilon m) + \frc{U^{\text{model}}}2 + \sum_{k=1}^\infty
    s_{2k}(2rm),
\eeq with
\begin{equation}\label{sell}
   s_{2k}(2rm) = \left.-\frac{d f_{2k}(2rm)}{dn}\right|_{n=1}.
\end{equation}
Here, we choose the universal model-dependent constant $U^{\text{model}}$ to be the same as in the
bulk expansion (\ref{bulkfinal}). This, as we explained in the introduction and in
subsection \ref{ssectnorm}, completely fixes the UV
behaviour of the entropy in the boundary cases, and in particular fixes the relation
between the non-universal short-distance cutoff $\varep$ and the correlation length.

\subsubsection{Next-to-leading order IR correction to the boundary entanglement entropy}

The series (\ref{boundaryfinal}) is of immediate use to provide a large-distance expansion
of the boundary entanglement entropy. In particular,
we may easily obtain the general form of the first correction to saturation
at large distances.
From (\ref{2p}) and (\ref{sell}) we see that the next-to-leading
order correction for $r$ large is given by
\begin{eqnarray}\label{s2}
s_2(2rm)&=&\frac{1}{4
    \pi }\int_{-\infty}^{\infty}d\theta\,
    R\left(\frac{i\pi}{2}-\theta\right)
    \left[\frac{d F_{2}^{\mathcal{T}|11}(-2\theta)}{dn}\right]_{n=1} e^{-2rm \cosh
    \theta}\nonumber\\
    &=&-\frac{1}{2}\int_{-\infty}^{\infty}d\theta\,
    \frac{R\left(\frac{i\pi}{2}-\theta\right)}{(\cosh \frac{\theta}{2})^2}
   \left[ \frac{F_{\text{min}}^{\mathcal{T}|11}(-2\theta)}{F_{\text{min}}^{\mathcal{T}|11}(i \pi)}\right]_{n=1} e^{-2rm \cosh
    \theta}.
\end{eqnarray}
This correction is much more model-dependent than the similar correction in the bulk case,
both through the boundary reflection matrix and through the minimal
form factors (which come from the scattering matrix); there is in fact
little to say concerning its general features.

In order to study the form of $s_2(2mr)$ in more detail and
investigate higher order corrections, we now
resort to the particular example of the Ising model.

\subsubsection{Boundary entanglement entropy of the Ising model} \label{sssectbis}

The expansion (\ref{boundaryfinal}) is not only an efficient large-distance expansion,
but also an exact representation. Hence, it can be used to analyse the short-distance behaviour of
the boundary entanglement entropy as well, by resumming all the terms of the series. In what follows,
we will reproduce the main arguments and results of such a study of in the case of
the Ising model, done in \cite{nexttonext}.

Let us recall the types of integrable boundary conditions that
have been found for the Ising model, whose two-particle $S$-matrix
is simply $-1$. A family that was studied in much detail in
\cite{Ghoshal:1993tm} is that corresponding to the presence of a
magnetic field that couples to the Ising spin field on the
boundary. The spin field is the order parameter, hence we are
looking at the scaling limit of the Ising spin chain in a
transverse magnetic field whose magnitude is slightly below its
critical value (ordered regime), and with a parallel magnetic
field on the boundary. The corresponding boundary reflection
matrix is given by
\begin{equation}\label{r}
    R(\theta)=-i \tanh\frac{1}{2}\left(\theta-\frac{i \pi}{2}\right)\frac{\kappa-i \sinh\theta}{\kappa+ i
    \sinh\theta},
\end{equation}
which includes, for special values of the parameter $\kappa$, the
following physically different types of integrable boundary
conditions,
\begin{itemize}
    \item \emph{{Free boundary condition}}: $\kappa=1$,
    \item \emph{{Fixed boundary condition}}: $\kappa = -\infty$,
    \item \emph{{Magnetic boundary conditions}} (interpolating between the previous
    two): $\kappa=1-\frac{h^2}{2m}$, where $h$ is a boundary
    magnetic field $0 < h < \infty$. The free boundary condition
    would then correspond to $h=0$ whereas the fixed boundary
    condition is equivalent to having a infinitely large magnetic
    field fixed at the boundary.
\end{itemize}
Boundary corrections to the expectation values of the energy and
disorder field in the Ising theory were computed using this
reflection matrix in \cite{Konik:1995ws}.

In the cases where $\kappa>0$, the reflection matrix has a pole on
the imaginary line on the physical sheet, $0<{\rm Im}(\theta)\leq
i\pi/2$. This implies that the boundary state expression
(\ref{bs}) is not correct. A modified expression exists
\cite{Ghoshal:1993tm}, but for simplicity,  we will not analyse
this case here. Hence, throughout we will consider $\kappa\leq 0$.
Note that the case $\kappa=0$ does not require modifications,
since the residue of the $R$-matrix vanishes at this point. At
$\kappa=0$, the bound state becomes weakly bound, and propagates
far into the bulk

The cases $\kappa>-1$, i.e. $h<h_c=2\sqrt{m}$, are also somewhat
special. In these cases, the R-matrix still has a pole on the
imaginary $\theta$ line, although not on the physical strip when
$\kappa\le0$. As noted in \cite{Ghoshal:1993tm}, the case
$\kappa=-1$ corresponds to a ``critical'' value of the magnetic
field, $h_c$, at which the reflection matrix happens to have a
third order zero at $\theta=0$.

We note also that the $R$ matrix at $\kappa=0$ is just
equal to the negative of the fixed-boundary condition $R$ matrix,
$\kappa=-\infty$, and that
\begin{eqnarray}
  R\left(\frac{i\pi}{2}-\theta\right) &=&
  -R\left(\frac{i\pi}{2}+\theta\right).\label{r1}
\end{eqnarray}

In order to make it clear that we are dealing with the Ising
model, we will explicitly write the $\kappa$ dependence in the
boundary corrections, $s_{2k}(t)\mapsto s_{2k}(t,\kappa)$.

Reproducing the main arguments of \cite{nexttonext}, we will analyse the
corrections $s_{2k}(t,\kappa)$ order by order. This will provide, by resummation
of these corrections, an exact evaluation of the constant $V(\kappa)$ appearing in (\ref{shla}).

As was mentioned in subsection \ref{ssectnorm}, we expect the
constant $V(\kappa)$ to depend only on the conformal boundary
condition that is reached as we bring the length $r$ to zero
(i.e.\ in the limit $r\ll m^{-1}$). There are two conformal
boundary conditions for the Ising model: zero magnetic field,
i.e.\ free boundary condition, and infinite magnetic field, i.e.\
fixed boundary condition (although, to be precise, the latter one
has two possibilities, for the two orientations of the boundary
magnetic field). The first one is a UV boundary condition, and the
second an IR boundary condition, from the point of view of the
boundary renormalisation group flow. Hence, for any fixed
$\kappa$, we expect that in the UV limit $mr\to0$, we obtain a
situation described by the UV, free boundary condition. On the
other hand, if the system is already at a fixed boundary condition
$\kappa=-\infty$, we expect that as $mr\to0$ we obtain a situation
described by the IR, fixed boundary condition. Hence, $V(\kappa)$
should be the same for any finite $\kappa$, but should take
another value for $\kappa=-\infty$.

We may wish to obtain the boundary entanglement entropy in
the situation where the bulk is conformal, but the boundary
itself is not conformal, with a magnetic boundary condition, that depends on the magnetic field.
In this case, we have to
send the bulk correlation length $m^{-1}$ to infinity while keeping both $r$ and the magnetic,
boundary correlation length $\eta$ finite, and of the same order.
The result will be a universal function of the ratio $r/\eta$.
This is the limit where $m^{-1}$ is greater that all other scales,
so that we must take simultaneously
$mr\to0,\, \kappa\to-\infty$ in an appropriate way. The way this limit is taken can be understood through
CFT considerations, but the four-particle calculation below
will make it clear that it is the product $\kappa mr$ that we need to keep fixed. In terms of the
magnetic field, this means that we need to keep the product $h^2r$ fixed
(both ways are equivalent in the limit $\kappa\to-\infty$). That is, the boundary
correlation
length $\eta$ is of the order of $h^{-2}$. We will not investigate this situation much further,
although we will discuss it below in relation with the four-particle contribution.
But we note that after taking this particular limit,
if we then make the boundary correlation length $\eta$ very large or very small compared
to $r$, we obtain a free/fixed conformal boundary condition. From this point of view,
it is clear that it is indeed the point $\kappa=-\infty$ that divides the regions in $\kappa$ where
the UV/IR conformal boundary condition is reached as we take $mr\to0$ keeping $\kappa$ fixed.

\subsubsection{Two-particle correction to the entropy in the Ising model}

In terms of the $R$-matrix (\ref{r}) and the minimal form factor
(\ref{is}), the correction (\ref{s2}) becomes
\begin{equation}\label{c1}
s_2(t,\kappa) = -\frac{1}{8}\int_{-\infty}^{\infty}  {d \theta}\,
\left( \frac{\kappa+\cosh\theta}
     {\kappa-\cosh\theta}\right)\left(\frac{\cosh\theta-1}{\cosh^2\theta}\right)e^{-t \cosh\theta}.
\end{equation}
We see that the correction $s_2(t,\kappa)$ is finite for all values of $t$,
including $t=0$ (zero distance). In particular, this correction is not expected to
contribute to the logarithmic term in the small-$mr$ behaviour of the boundary entanglement
entropy, (\ref{shla}). At $t=0$ it is possible to evaluate the
integral above explicitly:
\begin{equation}\label{neg}
 c_2(\kappa) := s_2(0,\kappa)=\frac{1}{4} - \frac{\pi }{8} + \frac{\pi }{4 \kappa}
 - \frac{{\sqrt{1 -\kappa}}\,\left( \pi  + 2\arcsin(\kappa) \right) }
   {4\kappa{\sqrt{1 + \kappa}}},
\end{equation}
and in particular
\begin{equation}\label{m10}
 c_2(-1)=\frac{10-3\pi}{8} \qquad \text{and}\qquad c_2(0)=
\frac{\pi-2}{8}.
\end{equation}
The number $c(\kappa)$ will contribute to the constant $V(\kappa)$ of (\ref{shla}).

\subsubsection{Four-particle correction to the entropy in the Ising model}

It is interesting to analyze in some detail the next correction to the
boundary entanglement entropy for the Ising model. For boundary
theories, this is the first correction for which the issue of
finding the right analytic continuation in $n$ plays again a
crucial role. As it turns out, that crucial role extends to all
other higher order corrections, which were explicitly found for
the Ising model \cite{nexttonext}. Also, the contribution $s_4(t,\kappa)$
gives us a lot of insight as to how the short-distance behaviour in
(\ref{shla}) can be recovered; in particular, how the two values of $V(\kappa)$
can come out, and how they are connected by the massless, non-conformal magnetic boundary
situation explained in paragraph \ref{sssectbis} above.

The four-particle boundary
correction is given by
\begin{equation}\label{s4}
    s_4(t,\kappa)=-\frac{1}{2}\left[\prod_{k=1}^{2}\int_{-\infty}^{\infty} \frac{d\theta_k}{4\pi}
    R\left(\frac{i\pi}{2}-\theta_{k}\right) e^{-t\cosh\theta_k}\right]\frac{d}{dn} \left[
    \sum_{i,j=1}^n \frc1{\bra{\cal T}\ket} F_{4}^{\mathcal{T}| ii jj}(-\theta_1, \theta_1, -\theta_2,
    \theta_2)
   \right]_{n=1},
\end{equation}
where
\begin{eqnarray}\label{sum}
     \sum_{i,j=1}^n F_{4}^{\mathcal{T}| iijj}(-\theta_1, \theta_1, -\theta_2,
    \theta_2)
    &=& n \sum_{j=0}^{n-1}F_{4}^{\mathcal{T}| 11 11 }(-\theta_1, \theta_1, (-\theta_2)^j,
    \theta_2^j).
\end{eqnarray}
Here we have used (\ref{tralala}) (in the case of the Ising model,
there is no need for the ordering of the particle indices),
as well as the notation $\theta^j = \theta+2\pi i j$.
Employing (\ref{f}) we find
\begin{eqnarray}
  && \frac{n}{\langle\mathcal{T}\rangle}\sum_{j=0}^{n-1}F_{4}^{\mathcal{T}| 11 11  }(-\theta_1, \theta_1, (-\theta_2)^j,
    \theta_2^j) \label{4}\\
    &&=n^2 K(2 \theta_1) K(2 \theta_2)+ n\sum_{j=0}^{n-1} \left(K(\theta_{12}^{-j})K(\theta_{12}^j)-K(\hat{\theta}_{12}^{-j})K(\hat{\theta}_{12}^j)
    \right),\nonumber
\end{eqnarray}
where the function $K(x)$ was introduced in (\ref{111}), and we use the ``hat'' notation introduced
after equation (\ref{el}). It is
simple to show that the $n^2 K(2 \theta_1) K(2 \theta_2)$ term
will give no contribution to the derivative at $n=1$  so that only
the terms in the sum remain. These terms will give a contribution,
since, employing (\ref{p2})-(\ref{p3}), they can actually be
rewritten as
\begin{equation}\label{re}
    -n\sum_{j=0}^{n-1}
    \left[|K(\theta_{12}-2 \pi i j)|^2 -|K(\hat{\theta}_{12}-2 \pi i j)|^2
    \right]\stackrel{\int}=  -2n\sum_{j=0}^{n-1}
    \left[|K(\theta_{12}-2 \pi i j)|^2
    \right],
\end{equation}
where the symbol $\stackrel{\int}=$ means equality up to
integration in $\theta_1$ and $\theta_2$ inside the integral
(\ref{s4}). We can then use the result (\ref{ident}) to show that
the sum (\ref{re}) is proportional to $\delta(\theta_{12})$. The
four-particle correction to the saturation value of the Ising
entanglement entropy is therefore
\begin{eqnarray}\label{s4t}
  s_4(t,\kappa) = \frac{1}{32}\int_{-\infty}^{\infty} d\theta
   R\left(\frac{i\pi}{2}-\theta\right)^2 e^{-2t \cosh \theta}=
   \frac{1}{32}\int_{-\infty}^{\infty} d\theta \left(\frac{\kappa+\cosh \theta}{\kappa - \cosh \theta}\right)^2
    \frac{1 - \cosh \theta}{1 + \cosh \theta}e^{-2t \cosh
   \theta}.\nonumber
\end{eqnarray}

The four-particle contribution is divergent as $t \rightarrow 0$. This
points to the fact that $s_4(t,\kappa)$ will contribute to the logarithmic
term in the short-distance behaviour in (\ref{shla}).
Technically, the reason for this is that the integrand of
(\ref{s4t}) is a function that tends to the value $-1$ as $\theta
\rightarrow \infty$ when $t=0$. Therefore the integral at $t=0$ is
divergent. In order to find the precise behaviour of the
correction as $t$ approaches 0, one can rewrite the integral above
as:
\begin{equation}\label{s4t2}
  s_4(t,\kappa) =
   \frac{1}{32}\int_{-\infty}^{\infty} d\theta \left[\left(\frac{\kappa+\cosh \theta}{\kappa - \cosh \theta}\right)^2
    \frac{1 - \cosh \theta}{1 + \cosh \theta}+1 \right]e^{-2t \cosh
   \theta} - \frac{1}{16} K_0 (2t).
\end{equation}
The behaviour of the Bessel function as $t$ goes to zero is
well-known,
\begin{equation}\label{bes}
    K_0(2t)= -\gamma-\log(t)+{O}(t^2\log t),
\end{equation}
where $\gamma=0.577216...$ is the Euler-Mascheroni constant.
Written in this form, the integral part is now a finite constant
at $t=0$, and we may define
\begin{eqnarray}\label{fc}
 c_4(\kappa)&=&
    \frac{1}{16}\int_{0}^{\infty} d\theta \left[\left(\frac{\kappa+\cosh \theta}{\kappa - \cosh \theta}\right)^2
    \frac{1 - \cosh \theta}{1 + \cosh \theta}+1 \right] \nonumber\\
    &=&-\frac{1}{8(1+\kappa)^2} \left[ 2\kappa - 3\kappa^2 -1 +
  \frac{\kappa\left(2\kappa -1\right) (\pi+2\arcsin\kappa)}
   {{\sqrt{1 - \kappa^2}}}\right],
\end{eqnarray}
with in particular
\begin{equation}
    c_4(-1)=\frac{23}{120}.
\end{equation}
Therefore
we have
\begin{equation}\label{uv4}
    s_4(t, \kappa) = \frac{1}{16}\log(t)+\frc{\gamma}{16} + c_4(\kappa)+o(1)
\end{equation}
In view of the short-distance
behaviour of the entanglement entropy (\ref{shla}), we expect that
the coefficients of the logarithmic divergencies at small $rm$
will add up to the finite number $1/12$ when all corrections are
considered. This has been proven numerically in \cite{nexttonext}.
Also, the constant $c_4(\kappa)$, like $c_2(\kappa)$ above, is a
part of the constant $V(\kappa)$ in (\ref{shla}); again, in
principle one should add up all such constants, for all
corrections, in order to obtain $V(\kappa)$.

Recall that $V(\kappa)$ takes only two possible values, one for $\kappa>-\infty$
and one for $\kappa=-\infty$. This implies that $V(-\infty)\neq \lim_{\kappa\to-\infty} V(\kappa)$.
It turns out that this inequality is true for $c_4(\kappa)$, and should hold as well as for infinitely many
constants $c_{2\ell}(\kappa)$ \cite{nexttonext}. Technically, we observe that although
the integral in (\ref{s4t2}) has the same value for
$\kappa=-\infty$ as for $\kappa=0$ for any $t>0$, we have
$\lim_{\kappa\to-\infty} c_4(\kappa) = 3/8$, different from $c_4(0) = 1/8$. The
explanation is that the limit $t\to0$ of the integral in
(\ref{s4t2}) as a function of $\kappa$ is not uniform. For all
values of $t>0$ we have $s_4(t,-\infty) = s_4(t,0)$, and there is
a maximum for $\kappa\in(-\infty,1)$ at a unique value
$\kappa=\kappa_0$. But as $t$ becomes smaller, the position of
this maximum shifts towards more negative values, until it reaches
$-\infty$ at $t=0$. There, if we take away the constant (as
function of $\kappa$) term $\frac{1}{16}\log(t)$ in order to make
the limit finite, the value of the maximum itself reaches
$\lim_{\kappa\to-\infty} c_4(\kappa)$. It is also possible to observe in the integral in
(\ref{fc}) that the symmetry between $\kappa=-\infty$ and
$\kappa=0$ is broken. Indeed, if $\kappa\to-\infty$, the
term in parenthesis can be approximated by 1 except for values of
$\theta$ where $\kappa+\cosh\theta\approx0$. These are very large
values of $\theta$, but they are not damped by any other factor,
hence the mistake in approximating by 1 is non-negligible for any
$\kappa$.

This means that the expansion (\ref{uv4}) is valid only
for $\kappa>-\infty$. For the case $\kappa=-\infty$, that is, the
fixed boundary condition, we have to consider the other order of
the limits: first $\kappa\to-\infty$, then $t\to0$. By the
symmetry between $\kappa=-\infty$ and $\kappa=0$, we define \beq
    c_4(-\infty) =: c_4(0) = \frc18
\eeq so that (\ref{uv4}) still holds in the case of a fixed
boundary condition, $\kappa=-\infty$.

Recall our discussion in paragraph \ref{sssectbis}, about the particular
massless limit whereby the boundary is still magnetic, not conformal. In this perspective,
it is instructive to obtain a more general small-$t$
expansion, where we take simultaneously $\kappa\to-\infty$. Let us
consider $t\to0$ with $-\kappa t = a$ fixed. We may use the change
of variable $s=\cosh\theta-1$ and write $s_4(t,\kappa)$ as \beq
    \frc1{16} \int_0^{\infty} ds \lt(\frc{k+1+s}{k-1-s}\rt)^2 \lt( -\frc{\sqrt{s}}{(s+2)^{3/2}} + \frc1{s+1}\rt) e^{-2t(s+1)}
    - \frc1{16} \int_0^{\infty} ds \lt(\frc{k+1+s}{k-1-s}\rt)^2 \frc{e^{-2t(s+1)}}{s+1}.
\eeq The first integral as a function of $\kappa$ has a uniform
limit as $t\to0$ on $\kappa\in[-\infty,0)$, so that we can
directly take $\kappa=-\infty$ and $t=0$; this gives $(2-\log
2)/16$. The second integral does not have a uniform limit, but it
can be evaluated explicitly:
\[
    \frc1{16(a+t)}\lt( 4a e^{-2t} - (a+t)  \Gamma(0,2t) e^{2t} - 8a (a+t) e^{2(a+t)} \Gamma(0,2(a+t))\rt)
\]
where $\Gamma(z,u)$ is the incomplete Gamma function,
$\int_u^\infty v^{z-1} e^{-v} dv$. The small-$t$ limit can then
easily be taken: \beq
    s_4(t,-a/t) = \frc1{16} \log(t) + \frac{\gamma}{16}+ c_4^\natural(a)+ O(t).
\eeq where \beq\label{c2nat}
    c_4^\natural(a) =  \frc{3}8 - \frc12 a e^{2a} \Gamma(0,2a) .
\eeq It is easy to see that $c_4^\natural(a)$ interpolates between
$\lim_{\kappa\to-\infty}c_4(\kappa)$ at $a=0$ to $c_4(-\infty)$ at
$a=\infty$.

\subsubsection{Higher order corrections to the entropy in the Ising model}

Many of the subtleties observed in the previous section for the
four-particle correction to the boundary entanglement entropy
generalise to higher orders, in particular the need to find the
analytic continuation in $n$ of our expressions and the
non-commutativity of the limits observed above. We do not wish to
go into the details of the computations involved, which are very
technical and cumbersome, even for the Ising model. However, let us
give an indication of how it works.

The main
complication arises from the fact that the pole- and
zero-structure of the form factors (in particular, the way some
poles and zeroes cancel each other) changes substantially as soon
as $n$ is allowed to take non-integer values. As a consequence,
the phenomenon observed in \cite{entropy}, and recalled above, of non-uniform
convergence of form factors as $n\to1$ is generalised to a
non-uniform convergence as $n\to k'$ for all positive integers
$k'\leq k$ for the $2k$-particle form factor. In order to obtain the correct
analytic continuation, the principle is that it is
obtained from the analytic function that describes form factor
contributions at values of $n$ large enough. This amounts to
evaluating first the contribution we would obtain analytically
continuing in $n$ around some integer and then adding the residues
of all the extra poles that are crossed by the integration
contours when bringing $n$ from infinity. This is naturally not the
only analytic continuation that is possible and therefore it
becomes quite crucial to find ways of checking it for consistency.
The exacy UV behaviour shown below provides good support, see
\cite{nexttonext} for a more extensive discussion.

The general formula for all higher-particle corrections found in
\cite{nexttonext} takes the form:
\begin{eqnarray}
    s_{4k}(t,\kappa) &=&\label{1ps}
    \frac{ \pi^2 (-1)^{k} }{2
    k}\left[\prod_{a=1}^{2k} \int_{-\infty}^{\infty} \frac{d \theta_{a}}{4\pi}\rt]
     \delta(\theta)\lt[
\left(%
\begin{array}{c}
 2k-2 \\
  k-1 \\
\end{array}%
\right) \prod\limits_{j=1}^{2k}  \frc{e^{-t \cosh \theta_{j}}
R\left(\frac{i\pi}{2}-\theta_{j}\right)}{
    \cosh\frac{\hat{\theta}_{j j+1}}{2}} \rt. \\ &&  \lt. -
 \sum_{j=1}^{k}\sum_{a=1}^{j-1} \sum_{q=\pm}
 \left(%
\begin{array}{c}
  2k-1 \\
  k-j \\
\end{array}%
\right)     (-1)^{j} \prod\limits_{i=1}^{2k}  \frc{e^{-t \cosh
(\theta_{i}+ q\frc{j-a}{2k}\pi i)}
 R\left(\frac{i\pi}{2}-(\theta_{i}+ q\frc{j-a}{2k}\pi i)\right)}{
  \cosh\lt(\frac{\h\theta_{i,i+1}}{2} + q\frc{j-a}{2k} \pi i\rt)} \rt] \no
\end{eqnarray}
and similarly
\begin{eqnarray}\label{1ps2}
    && s_{4k+2}(t,\kappa)  = \frac{ \pi^2 (-1)^{k} }{2
    k+1}\left[\prod_{a=1}^{2k+1} \int_{-\infty}^{\infty} \frac{d \theta_{a}}{4\pi}\rt]\delta(\theta) \\ &&\qquad \times
  \sum_{j=1}^{k}\sum_{a=1}^{j} \sum_{q=\pm}
 \left(%
\begin{array}{c}
  2k \\
  k-j \\
\end{array}%
\right)     (-1)^{j} q \prod\limits_{i=1}^{2k+1}  \frc{e^{-t \cosh
(\theta_{i}+ q\frc{j-a+1/2}{2k+1}\pi i)}
 R\left(\frac{i\pi}{2}-(\theta_{i}+ q\frc{j-a+1/2}{2k+1}\pi i)\right)}{
  \cosh\lt(\frac{\h\theta_{i,i+1}}{2} + q\frc{j-a+1/2}{2k+1} \pi i\rt)}, \no
\end{eqnarray}
both formulae hold for $k=1,2,3,\ldots$. All integrals involved
are absolutely convergent. Notice that for $k=1$ in (\ref{1ps}) we
recover the result (\ref{s4t}) as it should be. Setting $k=0$ in
(\ref{1ps2}) does not give (\ref{c1}), since $k=0$ is out of the
range of applicability of these formulae; the two-particle case
$s_2(t,\kappa)$ is a special case.

From these expressions it is also possible to extract the
divergent part as $t \rightarrow 0$, as done for the four-particle
contribution in the previous paragraph. This is achieved by simply
setting $t=0$ in all expressions, after subtracting 1 from the
product of $R$-matrices. We then obtain $s_{4k}(t,\kappa) \sim
c_{4k}(\kappa) + e_{2k}(t)$ and $s_{4k+2}(t,\kappa) \sim c_{4k+2}$ as
$t\to0$, for all $k=1,2,3,\ldots$, where $e_{2k}(t)$ are the
expansion coefficients (\ref{el}) in the expression for the bulk
entanglement entropy (see (\ref{bulkfinal}), and where \beqa
c_{4k}(\kappa) &=& \label{c2l}
    \frac{ \pi^2 (-1)^{k} }{2
    k}\left[\prod_{a=1}^{2k} \int_{-\infty}^{\infty} \frac{d \theta_{a}}{4\pi}\rt]
     \delta(\theta)\lt[
\left(%
\begin{array}{c}
 2k-2 \\
  k-1 \\
\end{array}%
\right) \frc{\prod\limits_{j=1}^{2k}
R\left(\frac{i\pi}{2}-\theta_{j}\right) - 1}{
    \prod\limits_{j=1}^{2k}  \cosh\frac{\hat{\theta}_{j j+1}}{2}} \rt. \\ &&  \lt. -
 \sum_{j=1}^{k}\sum_{a=1}^{j-1} \sum_{q=\pm}
 \left(%
\begin{array}{c}
  2k-1 \\
  k-j \\
\end{array}%
\right)     (-1)^{j}  \frc{ \prod\limits_{i=1}^{2k}
R\left(\frac{i\pi}{2}-(\theta_{i}+ q\frc{j-a}{2k}\pi i)\right) -
1}{
  \prod\limits_{i=1}^{2k} \cosh\lt(\frac{\h\theta_{i,i+1}}{2} + q\frc{j-a}{2k} \pi i\rt)} \rt] \no
\end{eqnarray}
and similarly
\begin{eqnarray}\label{c2lp1}
    c_{4k+2}(\kappa) &=& \frac{ \pi^2 (-1)^{k} }{2
    k+1}\left[\prod_{a=1}^{2k+1} \int_{-\infty}^{\infty} \frac{d \theta_{a}}{4\pi}\rt]\delta(\theta) \\ && \times
  \sum_{j=1}^{k}\sum_{a=1}^{j} \sum_{q=\pm}
 \left(%
\begin{array}{c}
  2k \\
  k-j \\
\end{array}%
\right)     (-1)^{j} q \prod\limits_{i=1}^{2k+1}  \frc{
 R\left(\frac{i\pi}{2}-(\theta_{i}+ q\frc{j-a+1/2}{2k+1}\pi i)\right)}{
  \cosh\lt(\frac{\h\theta_{i,i+1}}{2} + q\frc{j-a+1/2}{2k+1} \pi i\rt)}, \no
\end{eqnarray}
both formulae for $k=1,2,3,\ldots$.  The constant $V(\kappa)$ that
characterises the UV behaviour of the boundary entanglement
entropy as shown in (\ref{shla}) can now be obtained as follows:
\beqa
    S_A^{\rm boundary}(rm) &=& \frc1{12} \log(m\varep) + \frc{U^{\text{Ising}}}2 + \sum_{k=1}^{\infty} s_{2k}(2rm,\kappa) \n
        &\sim& \frc1{12} \log(m\varep) + \frc{U^{\text{Ising}}}2 +
            \sum_{k=1}^\infty c_{2k}(\kappa)
            + \frc12\lt(S_A^{\rm bulk}(2rm)- U^{\text{Ising}} - \frc1{6} \log(m\varep)\rt)  \n
        &\sim& -\frc1{12} \log(2r/\varep) +
            \sum_{k=1}^{\infty} c_{2k}(\kappa),
\eeqa hence
\begin{equation}\label{Vkappa}
    V(\kappa)=   \sum_{k=1}^{\infty} c_{2k}(\kappa).
\end{equation}
Naturally, in order for (\ref{Vkappa}) to be a correct
representation of $V(\kappa)$, the infinite sum over $k$ should
give a finite result. This is quite subtle, as form factor
expansions are expected to provide convergent series expansion for
finite distances, but not necessarily at zero distance. In
\cite{nexttonext}, exact evaluations of the first few coefficients
$c_{2k}(\kappa)$ for $\kappa=0,-1, -\infty$, and extrapolation to
higher $k$, gave strong indications that the series is indeed
convergent. More precisely we have managed to obtain closed
formulae for $c_{2k}(0)$ and $c_{2k}(-\infty)$:
\begin{equation}
    c_{4k}(0)=\frac{1}{8 k  (2k-1)},\qquad c_{4k+2}(0)=\frac{\pi}{2^{4k+1}}\left(
\begin{array}{c}
  2k-1 \\
  k-1 \\
\end{array}%
\right)^2-\frac{1}{4 (2k+1)},
\end{equation}
and $c_{4k}(-\infty) = c_{4k}(0)$ and $c_{4k+2}(-\infty) =
-c_{4k+2}(0)$. This then gives
\begin{equation}\label{v0}
    V(0)=\frac{\pi-2}{8}+\sum_{k=1}^\infty
    c_{4k}(0)=\log \sqrt{2},
\end{equation}
and \beq\label{Vminft}
    V(-\infty) = \frac{2-\pi}{8}+\sum_{k=1}^\infty
   ( c_{4k}(0)-c_{4k+2}(0))=0.
\eeq We have also computed the values of $c_{2k}(-1)$ for $k$ up
to 4 and found that $V(-1) \simeq 0.321966...$. This value is
compatible with $\log \sqrt{2}= 0.346574...$. Therefore, as
expected, in the UV limit the entropy only depends on whether the
magnetic field $h$ is finite (free boundary conditions) or
infinite (fixed boundary conditions). We can then summarise our
result as
 \beq\label{resVkappa}
    V(\kappa) = \lt\{\ba{ll} \log\sqrt{2} & (\kappa>-\infty) \\ 0 & (\kappa=-\infty). \ea \rt.
    \eeq

\subsubsection{Connection with the boundary entropy}

It is natural to interpret $V(\kappa)$ as a ``boundary
entanglement'': the contribution of the boundary to the
entanglement between the region $A$ and the rest. Naturally, for
fixed boundary condition, there should be no contribution at all,
since the boundary does not experience quantum fluctuations. Our
result (\ref{resVkappa}) shows that we have chosen the correct
large-distance normalisation to have $V(-\infty)=0$. On the other
hand, for free boundary conditions, the boundary fluctuates and
should participate to the entanglement. This is in agreement with
$V(\kappa>-\infty)=\log\sqrt{2}>0$.

In fact, it was shown in \cite{nexttonext} that we may connect $V(\kappa)$ to the {\em boundary entropy}
$s$, a quantity that essentially counts the number of degrees of
freedom pertaining to a boundary. This quantity is simply given by
$s=\log g$ where $g$ is the boundary degeneracy introduced by
Affleck and Ludwig \cite{Affleck:1991tk}. In particular for a bulk
CFT, they showed that $g=\bra0|\t{B}\ket$ where $|0\ket$ is the
bulk CFT ground state, and $|\t{B}\ket$ is a boundary
state of norm 1 in the bulk CFT Hilbert space (in particular, $s\leq 0$). It
is considering the boundary entropy that Friedan and Konechny
\cite{Friedan:2003yc} were able to provide a proof of the
``$g$-theorem'': that the $g$-function decreases in the RG flow
from UV to IR. For the Ising model, it turns out that $s=0$ in the free
boundary case, and $s=-\sqrt{2}$ in the fixed boundary case.

In order to understand the general relation between $V(\kappa)$ and $s$, we
must make use of the formula derived by Calabrese and Cardy \cite{Calabrese:2004eu}
connecting the boundary entropy to the entanglement entropy at criticality:
\beq\label{CC}
    s = S_A^{\rm boundary}(r)_{\rm critical} -
    \frc12 S_A^{\rm bulk}(2r)_{\rm critical}.
\eeq The result is independent of $r$, since both entanglement
entropies have the same logarithmic $r$-dependence. Naturally, the
universal entanglement entropy, as we emphasised in the
introduction, must be defined by subtracting a non-universal
constant, the logarithm of the correlation length. Hence, the
universal entanglement entropy is only defined up to addition of a
$r$-independent and boundary-independent constant. We provided in
the introduction precise ways of fixing this constant in the bulk
and boundary cases, by looking at asymptotics, with a massive
bulk. However, formula (\ref{CC}) holds if and only if both
entanglement entropies are evaluated in the same cut-off scheme.
For instance, both should be evaluated on the same lattice at
infinite correlation length, and taking the distance $r$ to be the
same number of lattice sites (this number tending to infinity).
Formula (\ref{CC}) is not of immediate use in the context of QFT,
because the requirement of a consistent cut-off scheme is a not a
universal requirement.

It turns out that the use of massive QFT solves this problem.
More precisely, we can make sure that we have the same cut-off schemes in calculating
$S_A^{\rm boundary}(r)_{\rm critical}$ and $S_A^{\rm bulk}(2r)_{\rm critical}$ by connecting
them to our universal definitions of the entanglement entropy in the bulk
and boundary massive cases. Recall that our definitions involve the inherently massive large-distance asymptotics.
More precisely, we can show that, in the Ising model,
\beq\label{calca}
    V(\kappa) = S_A^{\rm boundary}(r)_{\rm critical} - \frc12 S_A^{\rm bulk}(2r)_{\rm critical} + \log\sqrt{2}.
\eeq
The entanglement
entropies should be evaluated at the UV fixed point if $\kappa$ is
finite, and at the IR fixed point if $\kappa=-\infty$.

This implies that
\begin{equation}
    V(\kappa) = s - \log \sqrt{2}.\label{resVkappaising}
\end{equation}
In fact, it was argued in \cite{nexttonext} that in general the
shift $\log\sqrt{2}$ should be replaced by $-\log{\cal C}$ where
${\cal C}^2$ is the fraction of the ground state degeneracy broken
by the boundary condition for large enough $h$ (at the IR point).
In the Ising model, the ground state has a double degeneracy (the
two orientations of the spin), and a large magnetic field $h$
breaks it to 1, so that ${\cal C}^2 = 1/2$. This implies
\begin{equation}
    V(\kappa) = s - \log \mathcal{C}.\label{resVkappagen}
\end{equation}
Such an
``extra'' contribution $-\log\mathcal{C}$ to $s=\log g$ was also found in
\cite{Patrick} in calculations using Thermodynamic Bethe Ansatz
techniques (although it
is important to note that the Thermodynamic Bethe Ansatz used
there is fundamentally different from our approach based on form
factors). In \cite{Patrick} the flow of $g$ between critical points was
studied for several families of minimal Toda field theories with
ground state degeneracy $k$ and with a boundary completely
breaking it. The factor ${\cal C} = 1/\sqrt{k}$ was termed
``symmetry factor''. Interestingly, it was shown that at the
infrared point of these models, one gets $g={\cal C}$. Hence, for
massive models with spontaneously broken order-parameter symmetry,
and with an order-parameter boundary perturbation, one should find
$V(\kappa) = s-s_{IR}\geq 0$, where $s_{IR}=\log{\cal C}$ is the
infrared value of $s$. For the same order-parameter perturbation
both on the bulk and boundary, one should simply find
$V(\kappa)=s\leq0$. All these considerations should not depend on
integrability.

Let us now discuss briefly how we obtain (\ref{calca}), following \cite{nexttonext}.
Consider the
entanglement entropy $S_A^{\text{boundary}}(x_1,x_2)$ defined in subsection \ref{bent}, but
with a slightly different normalisation specified below; we will
denote it $\t{S}_A(x_1,x_2)$. We may uniquely fix the cutoff, for
instance by requiring the conformal normalisation,
$\t{S}_A(x_1,x_2)\sim \frc{c}3 \log(|x_2-x_1|/\varep) + o(1)$ as
$x_2\to x_1$. But this short distance behaviour is the bulk critical entanglement
entropy, so that we have chosen $\varep$ with $S_A^{\rm bulk}(r)_{\rm critical} = \frc{c}3
\log(r/\varep)$. Using the same object, hence with the same
lattice spacing, we can also define the critical entanglement
entropy in the boundary case, $S_A^{\rm boundary}(r)_{\rm
critical}$, following subsection \ref{bent}.
Note first that for both $x_2$ and $x_1$ far from the boundary and far
from each other, the
entanglement entropy saturates to some constant $-\frc{c}3
\log(m\varep) + \t{U}$ thanks to the presence of the mass.
This saturation is a sum of the contributions of the two boundary
points at $x_1$ and $x_2$, each one contributing $-\frc{c}6 \log(m\varep) +
\t{U}/2$. Now for $x_2$ alone being far from the boundary, we get
our usual boundary case, but with
an extra contribution of the boundary point $x_2$ at infinity.
Subtracting this contribution, we find $\t{S}_A(r)$ with $r$ the distance between $x_1$ and
the boundary: this is our usual boundary case,
with possibly a different cut-off definition. Taking the limit $rm\to0$,
we get $\t{S}_A(r) \sim S_A^{\rm boundary}(r)_{\rm critical} + o(1)$.

We must now relate $\t{S}_A(r)$ to $S_A^{\text{boundary}}(r)$. We know that for $r$ large,
the former saturates to $-\frc{c}6 \log(m\varep) +
\t{U}/2$, and the latter to $-\frc{c}6 \log(m\varep) +
U/2$. The difference between $\t{U}$ and $U$ is that the former really just counts
the contributions to the entanglement entropy around the boundary points of $A$
when $A$ is large, whereas the latter counts an additional entropy due to the ground
state degeneracy. Indeed, the constant $U$ occurs in the bulk, which should be seen
as a large-volume limit of a periodic space where both spin-up $|0,+\ket$ and spin-down
$|0,-\ket$ ground states are counted. The constant $\t{U}$ occurs when the system
has a boundary, where the ground state is fixed if the magnetic field is large enough.
Hence we have \beq\label{UtU}
    U = \t{U} + \log 2.
\eeq Then we find
$S_A^{\rm boundary}(r)_{\rm critical} =\frc{c}3 \log(r/\varep) +
V(\kappa)-\log\sqrt{2}$, which shows (\ref{calca}).

\begin{figure}
\begin{center}
\includegraphics[width=7.9cm,height=3.5cm]{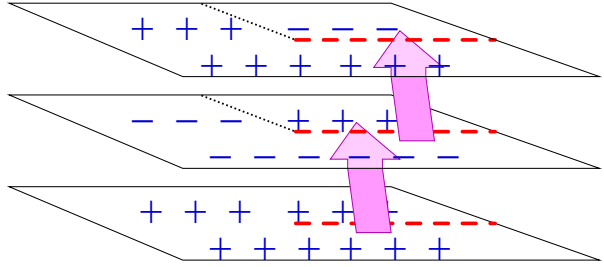}
\end{center}
 \caption{How a vacuum of the form
$\cdots \otimes |0,+\ket \otimes |0,-\ket \otimes |0,+\ket \otimes
\cdots$ becomes a two-particle state after going through a
branch-point twist field.} \label{figtp}
\end{figure}

There is a nice technical way of obtaining (\ref{UtU}), using
branch-point twist fields \cite{nexttonext}.
Allowing fluctuations amongst the two degenerate
ground states corresponds to choosing $|0\ket = ((|0,+\ket +
|0,-\ket)/\sqrt{2})^{\otimes n}$ in the $n$-copy model.
Our
branch-point twist field form factors assume the use of this
ground state. This can be
seen by recalling that particles in the Ising model are domain walls,
separations between up- and down-spin domains. In order to have
a non-zero matrix element of the branch-point twist field between
a state with a particle on one sheet and another particle on another sheet,
and the $n$-copy ground state, there must be components of the ground state
where one sheet is on $|0,+\ket$ and another sheet is on $|0,-\ket$.
See figure \ref{figtp}.
A completely up ground state $|0,+\ket^{\otimes n}$, for instance,
wouldn't admit such form factors. Since all configurations of an
even number of particles distributed amongst the $n$ sheets are
allowed, we must use the completely symmetric ground state. Then,
the constant $\t{U}$ is the large-distance limit of the bulk
entanglement entropy evaluated from $\bra 0,+|^{\otimes n} {\cal
T}(r_1) \t{\cal T}(r_2) |0,+\ket^{\otimes n}$, whereas $U$ is that
obtained from $\bra 0|\t {\cal T}(r_1) {\cal T}(r_2)|0\ket$ with
the symmetric ground state. The difference can be computed
explicitly. Note that matrix elements
\[
    \bra 0,\ep_1'|\otimes \cdots \otimes \bra 0,\ep_n'|  {\cal T}(r_1) \t{\cal T}(r_2) |0,\ep_1\ket\otimes \cdots \otimes |0,\ep_n\ket
\]
have zero large-distance limit unless $\ep_i=\ep_j=\ep_j'$ for all
$i,j$ (that is, all signs are the same), since otherwise domain
walls will have to propagate between the twist fields. Hence we
immediately find \beq
    \bra 0 | {\cal T}(r_1) \t{\cal T}(r_2)|0\ket \sim 2^{1-n} \bra 0,+|^{\otimes n} {\cal T}(r_1) \t{\cal T}(r_2) |0,+\ket^{\otimes n}
\eeq at large distances, so that, taking derivatives with respect
to $n$, we have \beq
    -\lt(\frc{d}{dn} \bra 0 |{\cal T}(r_1) \t{\cal T}(r_2)|0\ket \rt)_{n=1} \sim \log 2 - \lt(\frc{d}{dn}
    \bra 0,+|^{\otimes n} {\cal T}(r_1)  \t{\cal T}(r_2) |0,+\ket^{\otimes n} \rt)_{n=1}
\eeq which indeed gives (\ref{UtU}).

\sect{Conclusions and discussion}

 In this review we have summarised the fundamentals and
 main results of a program which was proposed in \cite{entropy}
 and developed further over a series of recent publications
 \cite{other,next,nexttonext}. This program allows for the computation of the entanglement entropy of
    a connected region $A$ a massive quantum one-dimensional system
    with respect to the remaining part of the system. To this aim, it takes full advantage of
    integrable quantum field theory techniques, as well as analytic scattering theory beyond integrability,
    by relating the bi-partite
    entanglement entropy to the correlation functions of a
    particular kind of local quantum fields: branch-point twist
    fields.

More precisely, we take as starting point  the well-known
``replica trick" and realise that, starting with a quantum
    integrable model and constructing a new model consisting of
    $n$ non-interacting copies of the original theory, a new local
    QFT is obtained which naturally possesses
    permutation symmetry. Associated to certain elements of this symmetry, two twist
    fields $\mathcal{T}$ and $\tilde{\mathcal{T}}$ exist, whose correlation functions are directly related
    to the bi-partite entanglement entropy. In general, there are as many branch-point twist fields in a correlation
    function as there are boundary points of the region $A$. We considered two cases,
    where the whole system does not have boundaries (bulk two-point functions), and where the system has one boundary and the region
    $A$ starts at this boundary (boundary one-point function), see Fig. \ref{chain}.
    Through the replica trick, the entropy is the derivative with respect to
    $n$ of the correlation function evaluated at
    $n=1$.

Since the fields $\mathcal{T}$ and $\tilde{\mathcal{T}}$ are local
fields of the $n$-copy theory, and the boundary state admits a
realization in terms of the scattering states of the bulk model,
the correlation functions (two- and one-point functions) above can
be computed by using the form factor expansion of massive quantum field theory.
They can be expressed as a sum for
different particle numbers over products of the form factors of
the fields involved. This gives a large-distance expansion:
computing the correlation function in the two-particle
approximation, for instance, gives the behaviour at large distances. This
expansion is in fact expected to converge rapidly, and the two-particle approximation
is often enough to describe the correlation
function up to relatively small distances.

The leading behaviour of the entropy as a function of the distance was
already well-known for very short and very large distances, both
for bulk and boundary models, before our work and is described by
the equations (\ref{shlabu}) and (\ref{shla}). For $r\ll m^{-1}$
($m$ being the mass of the lightest particle) that behaviour is
fixed by the properties of the underlying CFT which describes the
integrable QFT in the ultraviolet limit (in particular, its
central charge). Thus, the entropy can be computed explicitly by
using CFT techniques \cite{Calabrese:2004eu,Calabrese:2005in}. At
large separations $r\gg m^{-1}$ (in the infrared limit) the
entropy is known to saturate to a constant value.

One of the main results of our work has been to evaluate the first
correction to the bulk entropy in the infrared (large-distance) limit, providing
therefore a description of the behaviour of the entropy in the
intermediate region of values of $rm$. This correction is obtained
from the two-particle contributions to the form factor expansion.
In fact, the two-particle approximation provides both the
saturation value of the entropy, coming from the disconnected part
of the correlation function (the square of the vacuum expectation
value of the twist fields), and the exact first corrections up to
$O(e^{-3rm})$, coming from the two-particle contributions. The
exact value of the saturation in the Ising model (\ref{u}) was
computed in \cite{entropy} and showed there to be in good
agreement with previous numerical results \cite{Latorre2}.

The most surprising result of this analysis has been to establish
that the leading correction to the entropy at large $rm$ is in
fact a more universal quantity than expected, that is, it does not depend on the
particular scattering matrix of the model we started with, but
only on the spectrum of masses of the particles of the original
theory. This was shown to be true for all integrable QFTs (with or without
backscattering and/or bound states) \cite{entropy,other} and even for
non-integrable ones \cite{next}. It is quite remarkable that the
entropy should encode so explicitly crucial information about the
theory both in its UV regime (the central charge) and in the IR
regime (the number of light particles).

We have deduced this result from general arguments and checked it
explicitly for the Ising, sinh-Gordon models and sine-Gordon
models. The mathematical reason for this ``universal behaviour" is
clear, as the result is directly related to the presence of
kinematic poles in the two-particle form factors. Only form
factors having such poles do contribute to the final result for
the entropy, and their individual contributions turn out to be
independent of the scattering matrix. The presence of bound state poles does not
change this conclusion, neither does the loss of integrability.

For the quantum Ising model, we extended the main result above and
identified exact infinite-series formulae for the bi-partite
entanglement entropy both in the presence and absence of
boundaries \cite{nexttonext}.  In order to obtain our formulae, we
found closed expressions for all non-vanishing
    form factors of branch-point twist fields in the $n$-copy
    theory; we identified the correct analytic continuation in $n$ of
    the contributions of these form factors to correlation functions and evaluated their derivatives with respect to $n$; and
    we checked both the form factor formulae and their analytic continuation for consistency.
Since Wick's theorem applies, all form factors of the twist field
admit expressions in terms of a Pfaffian. From this, obtaining the right
analytic continuation of every contribution to the form factor
expansion of the twist-field boundary one-point function or bulk
two-point function is a highly non-trivial problem. We solved this problem
in \cite{nexttonext}, and verified it for consistency
in a very precise manner
by finding an explicit formula for the leading logarithmic
behaviour both of the two-point function of the twist field in the
bulk and of its derivative at $n=1$, and using a combination of analytical and
numerical computations.

The results just described for the Ising model have put us in the position to analyse
another quantity of interest in this context, that is the contribution to the free energy that can
be attributed exclusively to the presence of the boundary. This is
essentially the boundary entropy, the natural logarithm of the
boundary degeneracy or $g$-factor originally introduced by Affleck
and Ludwig in \cite{Affleck:1991tk}. In our analysis we have
computed the universal quantity $V(\kappa)$ which is closely
related to $g$ (where $\kappa$ is related to the boundary magnetic
field in the Ising model). We have defined $V(\kappa)$ in a universal, QFT way
as a certain $rm$ independent contribution to the boundary entanglement
entropy in the UV limit. We have found an exact formula for
$V(\kappa)$ and evaluated it at $\kappa=0,-\infty$ to
$V(0)=\log\sqrt{2}$ and $V(-\infty)=0$. We have also gathered
strong numerical evidence that $V(\kappa)$ is in fact constant and
equal to $V(0)$ for any finite values of $\kappa$. These two
values of $V(\kappa)$ would correspond to the two conformal
invariant boundary conditions that are known for the Ising model:
the free and fixed boundary conditions. The fact that $V(\kappa)$
is larger for free boundary conditions (finite magnetic field) and
that $V(0)-V(-\infty)=\log\sqrt{2}$ are properties which also hold
for $\log g$. However, it is known from Cardy and Lewellen's work
\cite{CardyLewellen} that $g_{\text{free}}=1$ and
$g_{\text{fixed}}=1/\sqrt{2}$, hence
\begin{equation}
    V(\kappa)-\log g=\log\sqrt{2}.
\end{equation}
We have identified the difference between these two quantities as
an IR contribution to the entanglement entropy coming from the
ground state degeneracy in the periodic case that is broken in the
boundary case. We have proposed a generalisation of this to more
general models with relevant boundary perturbations.

Although not discussed here in any detail, we have also found
interesting results for the IR behaviour of the entanglement
entropy in the boundary case. It is known that the bulk entropy
saturates for large distances. In particular, the fact that the entropy is an
increasing function of $rm$ follows from the ``strong
subadditivity theorem" and translation invariance, as proven in
\cite{Casini,Casini:2003ix}. Since translation invariance is
broken in the boundary theory, the entanglement entropy in this
case is not necessarily a monotonic function of $rm$. Indeed, we
find that, for a range of values of $\kappa$, it has a maximum for
some value of $rm$ before reaching its asymptotic value.
This range of values of $\kappa$
seems to start at $\kappa=-1$ which corresponds to a value of the
magnetic field for which, in a sense, the boundary becomes
``critical''.

There are many open problems related to our work and in general,
to the computation of the entanglement entropy in integrable QFT.
In the case of the bulk Ising model, it is known that the
entanglement entropy can be described via Painlev\'e transcendents
\cite{casini1,casini2}. It would be interesting to check the
consistency of this representation with our full form factor
expansion. In the boundary Ising case, it is known that the
one-point function of the order parameter has a Fredholm
determinant representation for any magnetic field, from which
differential equations can be derived \cite{Konik:1995ws}. It
would be interesting to see if similar formulae hold for the
branch-point twist field. In the general QFT case, the most
obvious problem is perhaps to extend the present boundary analysis to
theories other than the Ising model. We believe that this should
be possible to some extent but it is very unlikely that
re-summations can be done analytically for interacting models. Yet
an independent check of (\ref{resVkappagen}) in the more general
situation would be useful. It would also be interesting to apply
the form factor approach  to the computation of the entanglement
entropy of disconnected regions (see e.g.
\cite{Casini:2008wt,Calabrese:2009ez}), both for bulk and boundary
theories and also to extend the analysis of the present paper to
the finite temperature situation.

An additional result of our work has been to develop the form
factor program for branch-point twist fields in integrable models. As a consequence of
the particular exchange relations between these fields and the
fundamental fields of the theory, the form factor consistency
equations for branch-point twist fields are different from those
associated to standard local fields. In particular, the crossing
and kinematic residue equations are modified. Since here we have
only been concerned with the two-particle form factors of
branch-point twist fields, an interesting open problem remains,
namely to find closed solutions to the form factor equations for
arbitrary or at least higher particle numbers for models other
than the Ising theory. As explained in section \ref{higher}, we
expect these solutions to be given in terms of elementary
symmetric polynomials of the variables $e^{\theta_i /n}$. In fact,
the form factor program for more general twist fields is an aspect of
integrable QFT that still needs much more development.

\paragraph{Acknowledgments:}
We are grateful to J.~L. Cardy for his collaboration in the paper
\cite{entropy}, upon which this review is partly based. We also thank D. Bernard
for pointing out to us the paper \cite{BernardL92}.


\end{document}